\RequirePackage[l2tabu, orthodox]{nag}
\documentclass[a4paper, 11pt]{article}

\usepackage[top=1truein, bottom=1truein, left=1truein, right=1truein]{geometry}

\usepackage{type1cm}
\usepackage[T1]{fontenc}
\usepackage{lmodern}

\usepackage{amsmath}
\usepackage{amssymb}
\usepackage{bm}
\usepackage{mathtools}
\usepackage{textcomp}

\usepackage{blkarray}
\usepackage{multirow}
\usepackage{bigdelim}
\usepackage{booktabs}

\usepackage[bookmarksnumbered=true, hypertexnames=false, pdfdisplaydoctitle=true, setpagesize=false]{hyperref}
\usepackage{url}

\usepackage[capitalize, noabbrev]{cleveref}
\usepackage{autonum}

\usepackage{graphicx}
\usepackage{tikz}
\usetikzlibrary{patterns}
\usepackage{gnuplot-lua-tikz}

\usepackage{alphalph}
\usepackage{intcalc}
\usepackage{romannum}
\AtBeginDocument{\pagenumbering{arabic}}
\usepackage[subrefformat=parens]{subcaption}

\makeatother
\renewcommand\thesubfigure{%
  \alphalph{\intcalcAdd{\intcalcDiv{\expandafter\intcalcSub{\csname c@subfigure\endcsname}{1}}{3}}{1}}%
  -%
  \romannum{\intcalcAdd{\intcalcMod{\expandafter\intcalcSub{\csname c@subfigure\endcsname}{1}}{3}}{1}}%
}
\makeatletter

\usepackage{comment}

\usepackage[en-US]{datetime2}

\usepackage{okicmd}
\usepackage{okithm}

\numberwithin{equation}{section}

\title{%
  Improved Structural Methods\\%
  for Nonlinear Differential-Algebraic Equations\\%
  via Combinatorial Relaxation%
  \footnote{A preliminary version of this paper is to appear in Proceedings of the 44th International Symposium on Symbolic and Algebraic Computation (ISSAC 2019), Beijing, China, July 2019.}%
}

\author{Taihei Oki\footnote{%
  Department of Mathematical Informatics, Graduate School of Information Science and Technology, University of Tokyo, Tokyo 113-8656, Japan.
  E-mail: \href{mailto:taihei_oki@mist.i.u-tokyo.ac.jp}{\texttt{taihei\_oki@mist.i.u-tokyo.ac.jp}}
}}

\newcommand{\mykeywords}{differential-algebraic equations, index reduction, implicit function theorem, combinatorial relaxation, combinatorial scientific computing}

\makeatletter
\hypersetup{
  pdftitle={Improved Structural Methods for Nonlinear Differential-Algebraic Equations via Combinatorial Relaxation},
  pdfauthor={Taihei Oki},
  pdfsubject={},
  pdfkeywords={\mykeywords},
  pdfdisplaydoctitle=true
}
\makeatother

\newcommand{\setT}{\mathbb{T}}
\newcommand{\setI}{\mathbb{I}}
\newcommand{\ord}[2]{\sigma\prn{#1, #2}}
\renewcommand{\P}[1]{\mathrm{P}(#1)}
\newcommand{\D}[1]{\mathrm{D}(#1)}
\newcommand{\tdegdet}[1]{\hat{\delta}(#1)}
\newcommand\difop[1]{\mathrm{D}^{#1}}
\newcommand\sub[1]{\bar{#1}^{\mathrm{sub}}}
\newcommand\aug[1]{\bar{#1}^{\mathrm{aug}}}
\newcommand\LC[1]{\bar{#1}^{\mathrm{LC}}}

\newcommand\overmat[3]{%
  \BAmulticolumn{#1}{c}{%
    \overbrace{%
      \hphantom{%
        \begin{matrix}#2\end{matrix}%
      }%
    }^{#3}%
  }%
}

\begin{document}

\maketitle


\begin{abstract}
  Differential-algebraic equations (DAEs) are widely used for modeling of dynamical systems.
  In numerical analysis of DAEs, consistent initialization and index reduction are important preprocessing prior to numerical integration.
  Existing DAE solvers commonly adopt structural preprocessing methods based on combinatorial optimization.
  Unfortunately, the structural methods fail if the DAE has numerical or symbolic cancellations.
  For such DAEs, methods have been proposed to modify them to other DAEs to which the structural methods are applicable, based on the combinatorial relaxation technique.
  Existing modification methods, however, work only for a class of DAEs that are linear or close to linear.

  This paper presents two new modification methods for nonlinear DAEs: the substitution method and the augmentation method.
  Both methods are based on the combinatorial relaxation approach and are applicable to a large class of nonlinear DAEs.
  The substitution method symbolically solves equations for some derivatives based on the implicit function theorem and substitutes the solution back into the system.
  Instead of solving equations, the augmentation method modifies DAEs by appending new variables and equations.
  The augmentation method has advantages that the equation solving is not needed and the sparsity of DAEs is retained.
  It is shown in numerical experiments that both methods, especially the augmentation method, successfully modify high-index DAEs that the DAE solver in MATLAB cannot handle.
\end{abstract}

\begin{quote}
\end{quote}

\section{Introduction}
\label{sec:introduction}

Let $\setT \subseteq \setR$ be a nonempty open interval and $\Omega \subseteq \setR^{(l+1)n}$ a nonempty open set.
An $l$th-order \emph{differential-algebraic equation} (DAE) of size $n$ for $\funcdoms{x}{\setT}{\setR^n}$ is a differential equation in the form of
\begin{align}
  \label{eq:dae}
  F(t, x(t), \dot{x}(t), \ldots, x^{(l)}(t)) = 0,
\end{align}
where $\funcdoms{F}{\setT \times \Omega}{\setR^n}$ is a sufficiently smooth function.
DAEs have aspects of both ordinary differential equations (ODEs) $\dot{x}(t) = \phi(t, x(t))$ and algebraic equations $G(t, x(t)) = 0$.
DAEs are widely used for modeling dynamical systems such as mechanical systems, electrical circuits, and chemical reaction plants.

A fundamental and important problem in the study of DAEs is an \emph{initial value problem}, which is to find a smooth trajectory $\funcdoms{x}{\setT}{\setR^n}$ satisfying~\eqref{eq:dae} with the initial value condition
\begin{align} \label{eq:initial_value_condition}
  x(t^*) = x^*_{(0)},
  \quad
  \dot{x}(t^*) = x^*_{(1)},
  \quad\ldots,\quad
  x^{(l-1)}(t^*) = x^*_{(l-1)},
\end{align}
where $t^* \in \setT$ and $x_{(0)}^*, x_{(1)}^*, \ldots, x_{(l-1)}^* \in \setR^n$.
Unlike ODEs, an initial value problem for a DAE may not have a solution because the DAE can involve algebraic constraints, and the solution must satisfy not only the constraints but also their differentiations, called \emph{hidden constraints}.
While giving a consistent initial value of a DAE is an important process prior to numerical integration, this is known to be a non-trivial task~\cite{Brenan1996,Pantelides1988,Shampine2002}.

Another important preprocessing of the numerical simulation of DAEs is an \emph{index reduction}, which is a process of reducing the \emph{differentiation index}~\cite{Campbell1995a} of a DAE.
The differentiation index of a first-order DAE
\begin{align}
  \label{eq:1st-dae}
  F(t, x(t), \dot{x}(t)) = 0
\end{align}
is the minimum nonnegative integer $\nu$ such that the system of equations
\begin{align}
  F(t, x(t), \dot{x}(t)) = 0,
  \quad
  \dif{}{t} F(t, x(t), \dot{x}(t)) = 0,
  \quad \ldots, \quad
  \dif{^\nu}{t^\nu} F(t, x(t), \dot{x}(t)) = 0
\end{align}
can determine $\dot{x}$ as a continuous function of $t$ and $x$.
In other words, $\nu$ is the number of times one has to differentiate the DAE~\eqref{eq:1st-dae} to get an ODE.
Intuitively, the differentiation index represents how far the DAE is from ODEs.
The differentiation index of an $l$th-order DAE~\eqref{eq:dae} is defined as that of the first-order DAE obtained by replacing higher-order derivatives of $x$ with newly introduced variables.
It is commonly said to be difficult to numerically solve high $(\geq 2)$ index DAEs~\cite{Brenan1996,Hairer1996,Shampine2002}.
Therefore, it is important for accurate simulation of dynamical systems to convert a given DAE into a low $(\leq 1)$ index DAE.

Today, most simulation software packages for dynamical systems, such as Dymola, OpenModelica, MapleSim, and Simulink, are equipped with graph-based preprocessing methods, which we call \emph{structural methods}.
These methods were first presented by Pantelides~\cite{Pantelides1988} for the consistent initialization of DAEs.
This method was subsequently applied to an index reduction method by dummy derivative approach of Mattsson--S\"{o}derlind~\cite{Mattsson1993} (MS-method).
Pryce~\cite{Pryce2001} proposed a structural analysis method for DAEs, called the $\Sigma$-method, based on a variant of Pantelides' method.
These structural methods construct a bipartite graph from DAEs' structural information and solves an assignment problem on the bipartite graph.

These structural methods, however, do not work even for the following simple DAE
\begin{align} \label{eq:intro_dae}
  \left\{\begin{alignedat}{3}
    \dot{x}_1 +{} &\dot{x}_2 +       x_3 &{}= 0, \\
    \dot{x}_1 +{} &\dot{x}_2             &{}= 0, \\
                  &      x_2 + \dot{x}_3 &{}= 0.
  \end{alignedat}\right.
\end{align}
The $\Sigma$-method reports that the index is zero whereas it is indeed two.
This is because the method cannot detect the singularity of the coefficient matrix $\begin{pmatrix} 1 & 1 & 0 \\ 1 & 1 & 0 \\ 0 & 0 & 1 \end{pmatrix}$ of $\begin{pmatrix} \dot{x}_1 \\ \dot{x}_2 \\ \dot{x}_3 \end{pmatrix}$.
As this toy example shows, structural methods, which ignore numerical information, may fail on some DAEs due to numerical or symbolic cancellations.
In general, the structural methods work only if the associated Jacobian matrix, called the \emph{system Jacobian}, is nonsingular.

In order to overcome this issue for a first-order linear DAE
\begin{align} \label{eq:linear_DAE}
  A_1 \dot{x}(t) + A_0 x(t) = f(t)
\end{align}
with constant matrices $A_0, A_1 \in \setR^{n \times n}$ and a smooth function $\funcdoms{f}{\setT}{\setR^n}$, Wu et al.\@~\cite{Wu2013} presented a method to modify~\eqref{eq:linear_DAE} into an equivalent DAE having nonsingular system Jacobian, using the \emph{combinatorial relaxation} algorithm by Iwata~\cite{Iwata2003}. 
The combinatorial relaxation is a framework devised by Murota~\cite{Murota1995a} to solve linear algebraic problems by iteratively relaxing them into combinatorial optimization problems.
Another combinatorial relaxation method for linear DAEs whose coefficient matrices are \emph{mixed matrices} is given in~\cite{Iwata2018b}. 
A mixed matrix is a matrix consisting of accurate constants and inaccurate parameters.
Independently, Tan et al.~\cite{Tan2017} presented modification methods, called LC-method and ES-method, for nonlinear DAEs based on the same principle.
All the above methods iteratively replace an equation of DAEs by a linear combination of other equations or their derivatives.
These methods can deal only with DAEs close to linear DAEs; see \cref{sec:the_LC-method} for details.
In fact, one can make DAEs intractable just by changing the coordinate nonlinearly.

In this paper, we present two modification methods for nonlinear DAEs, which we call the \emph{substitution method} and the \emph{augmentation method}.
While the previous combinatorial relaxation methods~\cite{Iwata2003,Iwata2018b,Iwata2018a,Murota1995a,Tan2017,Wu2013} are designed only for a class of DAEs that is linear or close to linear, our methods are applicable to a much larger class of nonlinear DAEs.
The substitution method explicitly solves equations for some derivatives based on the implicit function theorem (IFT) and then substitutes the solution back into the system.
This can be seen as a generalization of solving linear equations in the LC-method.
To implement the substitution method, a routine to solve algebraic equations symbolically is needed.
The augmentation method is presented as a remedy for this drawback.
In order to avoid solving equations symbolically, the augmentation method introduces new variables and equations, which are copies of existing ones in the DAE system.
While the size of the modified DAE is increased, the augmentation method does not destroy the sparsity of DAEs.
We show in numerical experiments that both methods can modify high-index DAEs which cannot be dealt with by the standard DAE-solving library in MATLAB.
The experimental results also show that an equation-solving engine in MATLAB cannot obtain explicit functions in the substitution method depending on DAEs and on the selection of the values used in the method.
The augmentation method successfully serves as a remedy for this problem.

\paragraph{Related work.}
The substitution method repeatedly eliminates some derivatives in the DAE system.
In theory of DAEs and partial differential equations (PDEs), this approach is known as ``differential elimination'' or ``projection''~\cite{Gear1988,Qin2018,Reid2001}, especially for polynomial DAEs and PDEs.
Maple provides \texttt{rifsimp} function that simplifies polynomial PDEs based on the differential algebra and the Gr\"{o}bner basis~\cite{Maple}.
From practical dynamical systems, however, non-polynomial DAEs often appear.
Gear~\cite{Gear1988} described a na\"{i}ve index reduction method for nonlinear DAEs using a similar approach to the substitution method that iteratively eliminates derivatives using the IFT.
Gear's method appends differentiations of some equations in the DAE and thus the resultant DAE is overdetermined.
Our method is advantageous in this point since it returns DAEs having the same number of equations and variables.

Takamatsu--Iwata~\cite{Takamatsu2008} proposed an index reduction method which is also named as ``substitution method.''
Our substitution method is different from their substitution method in that their method deals with the first-order linear DAEs with constant coefficients based on combinatorial matrix theory, whereas our method is designed for fully nonlinear DAEs.

\paragraph{Organization.}
This paper is organized as follows.
\cref{sec:structural_methods_for_daes} summarizes structural methods for DAEs and analyzes failure reasons.
\cref{sec:dae_modification_via_combinatorial_relaxation} explains previous modification methods based on the combinatorial relaxation approach.
\cref{sec:substitution_method,sec:augmentation_method} describe the substitution method and the augmentation method, respectively.
\cref{sec:examples} illustrates two examples.
\cref{sec:numerical_experiments} shows results of numerical experiments.
Finally, \cref{sec:conclusion} concludes this paper.

\section{Structural Methods for DAEs}
\label{sec:structural_methods_for_daes}

\subsection{Preliminaries}

Structural methods for DAEs utilize information on which variable each equation depends.
We first introduce notations and a proposition to describe the structural methods.

Let $\setT \subseteq \setR$ be a nonempty open interval and $\Omega \subseteq \setR^{(l+1)n}$ a nonempty open set having coordinates $(x, \dot{x}, \ldots, x^{(l)})$, where $x^{(k)} = \prn[\Big]{x_j^{(k)}}_{j \in C} \in \setR^n$ for $k \in \set{0, 1, \ldots, l}$.
Here $C$ is a set of indices with $\card{C} = n$.
Note that each $x^{(k)}_j$ is regarded not as the $k$th-order derivative of some trajectory but as an independent variable here.
Let $\funcdoms{f}{\setT \times \Omega}{\setR}$ be a smooth function.
For $j \in C$ and $k \in \set{0, 1, \ldots, l}$, the function $f$ is said to \emph{depend on $x_j^{(k)}$} if the partial derivative $\pdif{f}{x_j^{(k)}}$ is not identically zero on the domain $\setT \times \Omega$ of $f$.
We denote the maximum nonnegative integer $k$ such that $f$ depends on $x_j^{(k)}$ by $\ord{f}{x_j}$.
If $f$ does not depend on $x_j^{(k)}$ for any $k$, we assign $\ord{f}{x_j} \defeq -\infty$ for convenience.

The \emph{derivative} $\dot{f}$ of $f$ with respect to $t$ is defined by
\begin{align} \label{def:derivative}
  \dot{f}(t, x, \dot{x}, \ldots, x^{(l+1)})
  \defeq
  \pdif{f}{t}(t, x, \dot{x}, \ldots, x^{(l)})
  +
  \sum_{k=0}^l \pdif{f}{x^{(k)}}(t, x, \dot{x}, \ldots, x^{(l)}) x^{(k+1)}.
\end{align}
For a nonnegative integer $d$, the $d$th-order derivative $f^{(d)}$ of $f$ is recursively defined by $f^{(0)} \defeq f$ and $f^{(d)} \defeq \dot{f}^{(d-1)}$ for $d \geq 1$.
It should be noted that the domain of $\dot{f}$ is not $\setT \times \Omega$ but $\setT \times \Omega \times \setR^n$ because $\dot{f}$ linearly depends on $x^{(l+1)}$.
Similarly, for a nonnegative integer $d$, we regard the domain of $f^{(d)}$ as $\setT \times \Omega^{(d)}$, where $\Omega^{(d)} \defeq \Omega \times \setR^{dn}$.

The following simple proposition plays an important role in structural methods for DAEs.

\begin{proposition}[{Griewank's lemma~\cite[Section~2.2]{Griewank1989}, \cite[Lemma~3.7]{Pryce2001}}] \label{lem:griewank}
  Let $\funcdoms{f}{\setT \times \Omega}{\setR}$ be a smooth function.
  For $j \in C$ and a nonnegative integer $d$, if $\ord{f}{x_j} \leq c$, then
  \begin{align} \label{eq:griewank}
    \pdif{f}{x_j^{(c)}}(t, x, \dot{x}, \ldots, x^{(l)})
    =
    \pdif{f^{(d)}}{x_j^{(c+d)}}(t, x, \dot{x}, \ldots, x^{(l+d)})
  \end{align}
  holds for all $(t, x, \dot{x}, \ldots, x^{(l+d)}) \in \setT \times \Omega^{(d)}$.
\end{proposition}

We sometimes regard the domain of $\pdif{f^{(d)}}{x_j^{(c+d)}}$ not as $\setT \times \Omega^{(d)}$ but as $\setT \times \Omega$ to simply write the equality~\eqref{eq:griewank} as $\pdif{f}{x_j^{(c)}} = \pdif{f^{(d)}}{x_j^{(c+d)}}$.
In addition, it follows from \cref{lem:griewank} that
\begin{align} \label{eq:ord_of_derivative}
  \ord{f^{(d)}}{x_j} = \ord{f}{x_j} + d
\end{align}
holds for $j \in C$ and a nonnegative integer $d$.

\subsection{Assignment Problem}

Pryce~\cite{Pryce2001} introduced an assignment problem for a reinterpretation of Pantelides' algorithm~\cite{Pantelides1988} as follows.

Consider a DAE~\eqref{eq:dae} of size $n$ with equation index set $R$ and variable index set $C$.
Let $G(F)$ denote the bipartite graph with vertex set $R \cup C$ and edge set
\begin{align}
  E(F) \defeq \set{(i, j) \in R \times C}[\ord{F_i}{x_j} > -\infty].
\end{align}
An edge subset $M \subseteq E(F)$ is called a \emph{matching} if the ends of edges in $M$ are disjoint.
A \emph{perfect matching} is a matching of size $n$.
We set the weight $c_e$ of an edge $e = (i, j) \in E(F)$ by $c_e = c_{i,j} = \ord{F_i}{x_j}$.

The assignment problem on $G(F)$ is the following problem $\P{F}$:
\Maximize[name=$\P{F}$]{%
  \sum_{e \in M} c_e%
}{%
  \text{$M \subseteq E(F)$ is a perfect matching on $G(F)$.}%
}%
The dual problem $\D{F}$ of $\P{F}$ is expressed as follows:
\Minimize[name=$\D{F}$]{%
  \sum_{j \in C} q_j - \sum_{i \in R} p_i%
}{%
  q_j - p_i \geq c_{i,j} & \prn{(i, j) \in E(F)}, \\
  p_i \in \setZ          & \prn{i \in R}, \\
  q_j \in \setZ          & \prn{j \in C}.
}%
It can be shown from the duality theorem that $\D{F}$ has an optimal solution if and only if $G(F)$ has a perfect matching.
Consider
\begin{align}
  \tdegdet{F} \defeq \text{the optimal value of the problem $\D{F}$},
\end{align}
which is equal to the optimal value of $\P{F}$ due to the strong duality.
If $\D{F}$ has no optimal solution, we assign $\tdegdet{F} \defeq -\infty$.
The problems $\P{F}$ and $\D{F}$ can be efficiently solved by the Hungarian method~\cite{Kuhn1955}. 

For a dual feasible solution $(p, q)$, a \emph{system Jacobian} $\funcdoms{D = (D_{i,j})_{i\in R, j\in C}}{\setT \times \Omega}{\setR^{n \times n}}$ of $F$ with respect to $(p, q)$ is a matrix defined by
\begin{align} \label{def:system_jacobian}
  D_{i,j}
  \defeq
  \pdif{F_i^{(p_i)}}{x_j^{(q_j)}}
  =
  \pdif{F_i}{x_j^{(q_j - p_i)}}
\end{align}
for each $i \in R$ and $j \in C$. 
The last equality in~\eqref{def:system_jacobian} for $(i, j)$ with $q_j - p_i \geq 0$ is due to \cref{lem:griewank}.
The equality also holds for $(i, j)$ with $q_j - p_i < 0$ by regarding $\pdif{F_i}{x_j^{(q_j - p_i)}}$ as an identically zero function.

Here we give a characterization of the optimality of $\D{F}$, which was originally given by Murota~\cite{Murota1995a} for linear DAEs with constant coefficients.
For a system Jacobian $D$, let $G^*(D)$ be the bipartite graph with vertex set $R \cup C$ and edge set
\begin{align} \label{def:edges}
  E^*(D) = \set{(i, j) \in R \times C}[\text{$D_{i, j}$ is not identically zero}].
\end{align}
The \emph{term rank} of $D$ is the maximum size of a matching in $G^*(D)$, and is denoted by $\trank D$.

\begin{proposition}[{\cite[Proposition~2.3]{Murota1995a}}] \label{prop:optimality}
  For a DAE~\eqref{eq:dae} of size $n$, let $D$ be a system Jacobian of the DAE with respect to a feasible solution $(p, q)$ of $\D{F}$.
  Then $(p, q)$ is optimal if and only if $\trank D = n$.
\end{proposition}

It is well-known that the term-rank of $D$ serves as a combinatorial upper bound on the rank of $D$.
Therefore, $\trank D = n$ is a necessary condition for the nonsingularity of $D$.

\subsection{Validity Condition for Structural Methods}

Pryce's $\Sigma$-method~\cite{Pryce2001} uses the assignment problem to determine a system of equations whose solution provides a consistent initial value.
The Mattsson--S\"{o}derlind method~\cite{Mattsson1993} (MS-method) reduces the index of DAEs in a structural way based on the dummy derivative approach.
The validity of these structural methods is established as follows.

\begin{theorem}[{\cite[Section~3.2]{Mattsson1993},~ \cite[Theorems~4.2,~5.2]{Pryce2001}}] \label{thm:validity_of_structural_methods}
  For a DAE~\eqref{eq:dae}, suppose that $\D{F}$ has an optimal solution $(p, q)$ and let $D$ be the system Jacobian of~\eqref{eq:dae} with respect to $(p, q)$.
  If there exists a consistent point $(t^*, X^*)$ of~\eqref{eq:dae} at which $D$ is nonsingular, then $(t^*, X^*)$ can be found by the $\Sigma$-method.
  In addition, the MS-method returns an equivalent DAE whose index is at most one around $(t^*, X^*)$.
\end{theorem}

In practice, the condition in \cref{thm:validity_of_structural_methods} is satisfied on many DAEs of real instances.
For example, Pryce~\cite{Pryce2001} showed that the $\Sigma$-method can be applied to any DAE which is of index zero, in standard canonical form, in Hessenberg form, a constrained mechanical system, or a triangular chain of systems for which the method works~\cite[Theorem~5.3]{Pryce2001}.
The structural methods succeed for seven instances out of nine DAE problems in the test set for IVP (initial value problem) solvers collected by Mazzia and Magherini~\cite{Mazzia2008}.

However, it is also true that the structural methods do not work for two DAEs in the test set, which model electrical circuits describing the behaviour of a transistor amplifier and a ring modulator.
In addition, it is reported~\cite{Iwata2018b,Scholz2018} that the structural methods fail for DAEs modeling simple RLC circuits.

Here we investigate how the structural methods fail.
From \cref{thm:validity_of_structural_methods}, these failures are classified into the following three scenarios.

\begin{enumerate}[label={(F\arabic*)}]
  \item The bipartite graph $G(F)$ has no perfect matching, or equivalently, the dual problem $\D{F}$ has no optimal solution.
  \item The system Jacobian $D$ with respect to an optimal solution of $\D{F}$ is not identically singular on $\setT \times \Omega$ but singular at all consistent points.
  \item $D$ is identically singular.
\end{enumerate}

Example DAEs of the failures are shown in the following.

\begin{example} \label{ex:F1}
  Consider the following DAE:
  \begin{align} \label{eq:example_F1}
    \left\{\begin{aligned}
      {x_1}^2 + (x_2 - 1)^2 &= 0, \\
      0 &= 0.
    \end{aligned}\right.
  \end{align}
  The DAE~\eqref{eq:example_F1} has a unique solution $x_1(t) = 0$ and $x_2(t) = 1$ for all $t \in \setR$.
  However, since the bipartite graph $G(F)$ corresponding to~\eqref{eq:example_F1} has no perfect matching, the structural methods cannot be applied to~\eqref{eq:example_F1} due to (F1).
  \qed
\end{example}

\begin{example} \label{ex:F2}
  Consider the following DAE:
  \begin{align} \label{eq:example_F2}
    \left\{\begin{aligned}
      {x_1}^2 &= 0, \\
      (x_2 - 1)^2 &= 0.
    \end{aligned}\right.
  \end{align}
  The solution of~\eqref{eq:example_F2} is the same as that of~\eqref{eq:example_F1}.
  
  We try to apply the $\Sigma$-method to~\eqref{eq:example_F2}.
  In Step~1, we find a dual optimal solution $p = (0, 0)$ and $q = (0, 0)$.
  The corresponding system Jacobian $D$ is
  \begin{align}
    D = \begin{pmatrix}
      2x_1 & 0 \\
      0 & 2(x_2 - 1)
    \end{pmatrix},
  \end{align}
  which is not identically singular on $\Omega = \set{(x_1, x_2)}[x_1, x_2 \in \setR]$.
  However, $D$ is singular at the unique consistent point $(0, 1)$ of~\eqref{eq:example_F2}.
  Hence~\eqref{eq:example_F2} does not satisfy the validity condition of the $\Sigma$-method (and the MS-method) due to (F2).
  \qed
\end{example}

\begin{example} \label{ex:F3}
  The structural methods cannot be applied to the DAE~\eqref{eq:intro_dae}.
  In fact, its system Jacobian $D$ corresponding to a dual optimal solution $p = (0, 0, 0)$ and $q = (1, 1, 1)$ is a singular constant matrix
  \begin{align}
    D = \begin{pmatrix}
      1 & 1 & 0 \\
      1 & 1 & 0 \\
      0 & 0 & 1
    \end{pmatrix}.
  \end{align}
  Thus the DAE~\eqref{eq:intro_dae} is in the case of (F3).
  \qed
\end{example}

The structural methods indeed fail for the aforementioned electrical network DAEs due to (F3).
In this paper, we focus on (F3).
It is also known that the nonsingularity of the system Jacobian is destroyed by a simple linear transformation of DAEs as follows.

\begin{example} \label{ex:linear_transform}
  Let $F(t, x, \dot{x}, \ldots, x^{(l)}) = 0$ be a DAE and $D$ the system Jacobian with respect to a dual optimal solution $(p, q)$.
  Suppose $p_{i_1} \neq p_{i_2}$ for some $i_1, i_2 \in R$.
  Take a ``generic'' matrix $A \in \setR^{n \times n}$, that is, each entry in $A$ is chosen at random.
  Then $F(t, x, \dot{x}, \ldots, x^{(l)}) = 0$ and $AF(t, x, \dot{x}, \ldots, x^{(l)}) = 0$ are equivalent DAEs since $A$ is nonsingular (with probability one), whereas $AF(t, x, \dot{x}, \ldots, x^{(l)}) = 0$ meets (F3) as we explain below.
  
  In fact, from the genericity of $A$, the associated graph $G(AF)$ is the complete bipartite graph with edge weight $\displaystyle c'_{h,j} \defeq \max_{i \in R} \ord{F_i}{x_j}$ for $(h, j) \in E(AF)$.
  Thus an optimal solution $(p', q')$ of $\D{AF}$ is $p' = (0, \ldots, 0)$ and $\displaystyle q' = \prn[\Big]{\max_{i \in R} \ord{F_i}{x_j}}_{j \in C}$.
  It is easy to see that the system Jacobian $D'$ of $AF(t, x, \dot{x}, \ldots, x^{(l)}) = 0$ with respect to $(p', q')$ is given by $D' = A\tilde{D}$, where $\tilde{D}$ is a matrix defined by $\tilde{D}_{i,j} \defeq D_{i,j}$ if $\ord{F_i}{x_j} = q'_j$ and $\tilde{D}_{i,j} \defeq 0$ otherwise for $i \in R$ and $j \in C$.
  Here from the assumption $p_{i_1} \neq p_{i_2}$, there is a row of zeros in $\tilde{D}$, and thus $D'$ is identically singular.
  \qed
\end{example}

The failure (F3) is attributed to the fact that the structural methods use only combinatorial information and ignore numerical and symbolic information of DAEs assuming that nonzero entries in Jacobian matrices are generic.
Then numerical or symbolic cancellations inherent in the DAEs make the system Jacobian identically singular.
\section{DAE Modification via Combinatorial Relaxation}
\label{sec:dae_modification_via_combinatorial_relaxation}

\subsection{Combinatorial Relaxation}
\label{sec:combinatorial_relaxation}

The method of Wu et al.~\cite{Wu2013} modifies a given first-order linear DAE~\eqref{eq:linear_DAE} with constant coefficients into an equivalent linear DAE without (F3), i.e., the system Jacobian is not identically singular.
This method relies on the combinatorial relaxation algorithm of Iwata~\cite{Iwata2003}, and all other modification methods are also based on the combinatorial relaxation approach.
The combinatorial relaxation method consists of the following three phases~\cite{Iwata2003,Murota1995a,Tan2017,Wu2013}.

\paragraph{Combinatorial Relaxation}

\begin{enumerate}[label={\textbf{Phase \arabic*.}}]
  \item
    Compute an optimal solution $(p, q)$ of $\D{F}$.
    If $\D{F}$ has no optimal solution, the algorithm terminates with failure.
  \item
    If the system Jacobian $D$ with respect to $(p, q)$ is not identically singular, return the DAE $F = 0$ and halt.
  \item
    Modify the DAE $F = 0$ into an equivalent DAE $\bar{F} = 0$ such that $\tdegdet{\bar{F}} \leq \tdegdet{F} - 1$.
    Go back to Phase~1.
\end{enumerate}

Since $\D{F}$ has an optimal solution if and only if $\tdegdet{F} \geq 0$, the above process ends in at most $\tdegdet{F} \leq ln$ iterations.
Therefore, given a DAE with (F3), the combinatorial relaxation method returns an equivalent DAE without (F3) (or with (F1) if the method has failed in Phase~1).

A non-trivial part of the combinatorial relaxation method is only Phase~3, which modifies DAEs to decrease the value of $\hat{\delta}$.
Iwata's combinatorial relaxation algorithm modifies first-order linear DAEs with constant coefficients using \emph{strict equivalence transformations}, which multiply nonsingular constant matrix to equations and variables.
A combinatorial relaxation method in~\cite{Iwata2018b} for linear DAEs with mixed matrices employs \emph{unimodular transformations} here.
The unimodular transformation is a sequence of trivial equivalent transformations of DAEs that add an equation (or its derivative) to another equation. 
Iwata--Takamatsu's index reduction algorithm~\cite{Iwata2018a} for first-order linear DAEs with constant coefficients is also based on the combinatorial relaxation and modifies DAEs using unimodular transformations.

\subsection{The LC-method}
\label{sec:the_LC-method}

The LC-method of Tan et al.~\cite{Tan2017} can be regarded as a nonlinear generalization of the method of Wu et al~\cite{Wu2013}, where the difference is only the modification method in Phase~3.
The modification method of the LC-method is summarized as follows.

Suppose that we have a DAE~\eqref{eq:dae} and its dual optimal solution $(p, q)$ such that the system Jacobian $D$ with respect to $(p, q)$ is identically singular.
First, we find a nonzero vector $u(t, x, \dot{x}, \ldots) = (u_i)_{i \in R}$ in the cokernel of $D$, namely, $u$ is a row vector such that $uD$ is identically zero.
Let $\supp u$ denote the support of $u$, i.e.,
\begin{align} \label{def:supp}
  \supp u \defeq \set{i \in R}[\text{$u_i$ is not identically zero}].
\end{align}
Take $r \in \supp u$ such that $p_r \leq p_i$ for all $i \in \supp u$ and put $I \defeq \supp u \setminus \set{r}$.
Then we replace the $r$-th equation $F_r = 0$ of the DAE by $\LC{F}_r = 0$, where
\begin{align} \label{eq:LC_F_r}
  \LC{F}_r \defeq u_r F_r + \sum_{i \in I} u_i F_i^{(p_i - p_r)}.
\end{align}

It is shown that this modification decreases the value of $\hat{\delta}$ if
\begin{align} \label{cond:LC}
  \ord{u_i}{x_j} < q_j - p_r
\end{align}
for all $i \in R$ and $j \in C$~\cite[Theorem~4.1]{Tan2017}.
Intuitively, the condition~\eqref{cond:LC} means that the highest-order derivatives appear linearly in DAEs.
For (time-varying) linear DAEs,~\eqref{cond:LC} trivially holds since $\ord{u_i}{x_j} = -\infty$ for all $i, j$.

However, there still exist DAEs to which the LC-method cannot be applied.
For example, the following DAE
\begin{align} \label{dae:LC_failure}
  \begin{cases}
    \dot{x}_1 \dot{x}_2 - 2 \cos^2 t = 0, \\
    {{}\dot{x}_1}^2 {{}\dot{x}_2}^2 + x_1 + x_2 - 4 \cos^4 t - 3 \sin t - 2 = 0
  \end{cases}
\end{align}
given in~\cite[Section~5.3]{Tan2017} cannot be dealt with by the LC-method.
While \cite{Tan2017} also presents another modification method called the ES-method, it is also inapplicable to~\eqref{dae:LC_failure}.
Indeed, the following example, which is a nonlinear generalization of \cref{ex:linear_transform}, demonstrates that one can convert many DAEs to other DAEs not satisfying~\eqref{cond:LC} by nonlinearly changing the coordinate of the codomain of $F$.

\begin{example} \label{ex:nonlinear_coordinate_change}
  Let $F(t, x, \dot{x}, \ldots, x^{(l)}) = 0$ be a DAE and $D$ the system Jacobian with respect to a dual optimal solution $(p, q)$.
  Suppose $p_{i_1} \neq p_{i_2}$ for some $i_1, i_2 \in R$ as in \cref{ex:linear_transform}.
  Let $\funcdoms{\psi = (\psi_h)_{h \in R'}}{\setR^n}{\setR^n}$ be a ``generic'' nonlinear diffeomorphism such that $\psi(w) = 0$ if and only if $w = 0$.
  Then $F(t, x, \dot{x}, \ldots, x^{(l)}) = 0$ is equivalent to $\psi(F(t, x, \dot{x}, \ldots, x^{(l)})) = 0$, whereas the latter DAE meets (F3) but cannot be handled by the LC-method since the highest-order derivatives appear nonlinearly.
  
  More formally, this is shown as follows.
  From the genericity assumption on $\psi$, it holds $\displaystyle \ord{\psi_h(F)}{x_j} = \max_{i \in R} \ord{F_i}{x_j}$ for each $h \in R'$ and $j \in C$.
  Then as in \cref{ex:linear_transform}, $G(\psi(F))$ is the complete bipartite graph, and a dual optimal solution is given by $p' = (0, \ldots, 0)$, $\displaystyle q' = \prn[\Big]{\max_{i \in R} \ord{F_i}{x_j}}_{j \in C}$.
  Let $\tilde{D}$ be a matrix defined by $\tilde{D}_{i,j} \defeq D_{i,j}$ if $\ord{F_i}{x_j} = q'_j$ and $\tilde{D}_{i,j} \defeq 0$ otherwise for $i \in R$ and $j \in C$.
  Then the $(h,j)$-th entry in the system Jacobian $D'$ of $\psi(F(t, x, \dot{x}, \ldots, x^{(l)})) = 0$ with respect to $(p', q')$ is
  \begin{align}
    D'_{h,j}
    &= \pdif{\psi_h(F(t, x, \dot{x}, \ldots, x^{(l)}))}{x_j^{(q'_j)}} \\
    &= \sum_{i \in R} \pdif{\psi_h}{w_i}(F(t, x, \dot{x}, \ldots, x^{(l)})) \pdif{F_i}{x_j^{(q'_j)}}(t, x, \dot{x}, \ldots, x^{(l)}) \\
    &= \sum_{i \in R} \pdif{\psi_h}{w_i}(F(t, x, \dot{x}, \ldots, x^{(l)})) \tilde{D}_{i,j}.
  \end{align}
  Therefore, it holds $D' = \dif{\psi}{w}(F(t, x, \dot{x}, \ldots, x^{(l)})) \tilde{D}$, where $\dif{\psi}{w}$ is the Jacobian matrix of $\psi$.
  Now there is a row of zeros in $\tilde{D}$ from the assumption $p_{i_1} \neq p_{i_2}$, and thus $D'$ is identically singular.
  In addition, a cokernel vector $u$ of $D'$ corresponds to a cokernel vector $v$ of $\tilde{D}$ by $u = \dif{\psi}{w}(F(t, x, \dot{x}, \ldots, x^{(l)})) v$.
  Since each entry in $\dif{\psi}{w}(F(t, x, \dot{x}, \ldots, x^{(l)}))$ depends on $x_j^{(q_j')}$ by the nonlinearity and genericity assumptions on $\psi$, an entry $u_h$ in $u$ depends on $x_j^{(q_j')}$ for each $h \in R'$ and $j \in C$ if $u \neq 0$.
  This means that the DAE $\psi(F(t, x, \dot{x}, \ldots, x^{(l)})) = 0$ does not fulfill the validity condition~\eqref{cond:LC} of the LC-method.
  \qed
\end{example}

The claim in \cref{ex:nonlinear_coordinate_change} implies that \eqref{cond:LC} holds only if we have a special coordinate of the codomain space of $F$.
Therefore, from a geometrical point of view, it is natural and important to devise a modification method for such ``heavily nonlinear'' DAEs.

\section{Substitution Method}
\label{sec:substitution_method}

\subsection{Outline of Method}
\label{sec:substitution_method_description}

In this section, we describe a new modification method for nonlinear DAEs, called the substitution method.
This method is used in Phase~3 of the combinatorial relaxation framework.

Let $\setT \subseteq \setR$ be a nonempty open interval and $\Omega \subseteq \setR^{(l+1)n}$ a nonempty open set.
The input of the substitution method is a DAE~\eqref{eq:dae} of size $n$ with real analytic function $\funcdoms{F}{\setT \times \Omega}{\setR^n}$ such that
\begin{enumerate}[label={(I\arabic*)}]
  \item $G(F)$ has a perfect matching,
  \item for any square submatrix $D[I, J]$ of the system Jacobian $D$ with respect to a dual optimal solution, if $D[I, J]$ is not identically singular on $\setT \times \Omega$, then there exists a consistent point of~\eqref{eq:dae} at which $D[I, J]$ is nonsingular, and
  \item $D$ is identically singular.
\end{enumerate}
The smoothness assumption on $F$ is needed to avoid technical difficulties.
We remark that (I2) is just a part of a sufficient condition for which the substitution method works, and it suffices in practice to check the condition only for a few submatrices of $D$ that are needed in the method.

The substitution method modifies the DAE~\eqref{eq:dae} into another DAE
\begin{align} \label{eq:sub_output_dae}
  \sub{F}(t, x, \dot{x}, \ldots, x^{(l+\kappa)}) = 0
\end{align}
of size $n$ such that
\begin{enumerate}[label={(S\arabic*)}]
  \item $\sub{F}$ is a real analytic function defined on a nonempty open subset $\sub{\setT} \times \sub{\Omega} \subseteq \setT \times \Omega^{(\kappa)}$ with $\kappa \leq ln$,
  \item the resulting DAE~\eqref{eq:sub_output_dae} is locally equivalent to the input DAE~\eqref{eq:dae}, and
  \item $\tdegdet{\sub{F}} \leq \tdegdet{F} - 1$.
\end{enumerate}
See \cref{lem:S2} for the precise meaning of ``locally equivalent'' in (S2).

We first introduce notations needed to describe the method.
Let $R$ and $C$ be the equation index set and the variable index set of the DAE~\eqref{eq:dae}, respectively.
For $I \subseteq R$, let $F_I$ denote a ``subvector'' $(F_i)_{i \in I}$ of $F$ indexed by $I$.
Similarly, for $J \subseteq C$, let $x_J$ denote a subvector $(x_j)_{j \in J}$ of $x$ indexed by $J$.
Let $p$ and $q$ be the vectors of variables in $\D{F}$.
In addition, we use the following notations
\begin{align}
  F_I^{(p)} \defeq \prn[\bigg]{F_i^{(p_i)}}_{i \in I},
  \quad
  x_J^{(q)} \defeq \prn[\bigg]{x_j^{(q_j)}}_{j \in J},
  \quad
  \pdif{F_I^{(p)}}{x_J^{(q)}}
  \defeq
  \prn{\pdif{F_i^{(p_i)}}{x_j^{(q_j)}}}_{i \in I, j \in J}
\end{align}
for $I \subseteq R$ and $J \subseteq C$.

Here we start to describe the method.
Let $D$ be the system Jacobian of~\eqref{eq:dae} with respect to an optimal solution $(p, q)$ of $\D{F}$ and suppose that $D$ is identically singular.
We regard $D$ as a matrix over the quotient field $\setF$ of the ring of real analytic functions on $\setT \times \Omega$.
The substitution method first finds $r \in R$, $I \subseteq R \setminus \set{r}$ and $J \subseteq C$ with $\card{I} = \card{J} \eqdef m$ such that
\begin{enumerate}[label={(C\arabic*)}]
  \item $D[I, J]$ is nonsingular,
  \item $\rank D[I \cup \set{r}, C] = m$, and
  \item $p_{r} \leq p_i$ for $i \in I$.
\end{enumerate}
Here, both the nonsingularity in (C1) and the rank in (C2) are in the sense of those of matrices over $\setF$.
Namely, these conditions can be rewritten as
\begin{enumerate}[label={($\text{C\arabic*}^*$)}]
  \item $D[I, J]$ is not identically singular, and
  \item the maximum size of a submatrix in $D[I \cup \set{r}, C]$ that is not identically singular is $m$.
\end{enumerate}
The existence of $(r, I, J)$ satisfying (C1)--(C3) is guaranteed through the algorithm explained in \cref{sec:algorithm_for_finding_rIJ}.

Let $(r, I, J)$ be a triple satisfying the conditions (C1)--(C3).
Define $S = R \setminus (I \cup \set{r})$ and $T = C \setminus J$.
Then the DAE~\eqref{eq:dae} is divided into three subsystems as follows:
\begin{align} \label{eq:fHK}
  \left\{ \begin{aligned}
    F_r(t, x, \dot{x}, \ldots, x^{(l)}) &= 0,\\
    F_I(t, x, \dot{x}, \ldots, x^{(l)}) &= 0, \\
    F_S(t, x, \dot{x}, \ldots, x^{(l)}) &= 0.
  \end{aligned} \right.
\end{align}
The system Jacobian $D$ with respect to $(p, q)$ forms a block matrix as follows:
\begin{align}
  D = \begin{blockarray}{ccc}
    & J & T \\
    \begin{block}{c(cc)}
      \set{r} & \pdif{F_r^{(p_r)}}{x_J^{(q)}} & \pdif{F_r^{(p_r)}}{x_{T}^{(q)}} \\
      I       & \pdif{F_I^{(p)}}{x_J^{(q)}} & \pdif{F_I^{(p)}}{x_{T}^{(q)}} \\
      S       & \pdif{F_{S}^{(p)}}{x_J^{(q)}} & \pdif{F_{S}^{(p)}}{x_{T}^{(q)}} \\
    \end{block}
  \end{blockarray}\,.
\end{align}

By the condition (C3) and \cref{lem:griewank}, it holds that
\begin{align}
  \pdif{F_I^{(p)}}{x_J^{(q)}}
  =
  \prn{\pdif{F_i^{(p_i)}}{x_j^{(q_j)}}}_{i \in I, j \in J}
  =
  \prn{\pdif{F_i^{(p_i - p_r)}}{x_j^{(q_j - p_r)}}}_{i \in I, j \in J}
  =
  \pdif{F_I^{(p - p_r \onevec)}}{x_J^{(q - p_r \onevec)}},
\end{align}
where $\onevec$ is the vector of ones with appropriate dimension.
In addition, from the condition (C1), the submatrix $D[I, J] = \pdif{F_I^{(p)}}{x_J^{(q)}} = \pdif{F_I^{(p - p_r \onevec)}}{x_J^{(q - p_r \onevec)}}$ is not identically singular on $\setT \times \Omega$.
Therefore, by (I2), there exists a point $(\hat{t}, \hat{X}) \in \setT \times \Omega^{(\kappa)}$ such that $F_I^{(p - p_r\onevec)} (\hat{t}, \hat{X}) = 0$ and $\pdif{F_I^{(p - p_r \onevec)}}{x^{(q - p_r \onevec)}}(\hat{t}, \hat{X})$ is nonsingular, where
\begin{align} \label{def:kappa}
  \displaystyle \kappa \defeq \max_{i \in I} p_i - p_r.
\end{align}
Then via the IFT, we can solve an equation
\begin{align} \label{eq:F_I_diff}
  F_I^{(p - p_r\onevec)}(t, x, \dot{x}, \ldots, x^{(l+\kappa)}) = 0
\end{align}
for $x_J^{(q - p_r\onevec)}$ as
\begin{align} \label{eq:phi}
  x_J^{(q - p_r\onevec)} = \phi(t, x, \dot{x}, \ldots, x^{(l+\kappa)}),
\end{align}
where $\phi$ is a function that does not depend on $x_J^{(q - p_r\onevec)}$.
See \cref{sec:application_of_implicit_function_theorem} for a rigorous description of this part.

Finally, we substitute the right-hand side of~\eqref{eq:phi} into $x_J^{(q - p_r\onevec)}$ in the first equation $F_r = 0$ of~\eqref{eq:fHK}.
The modified DAE~\eqref{eq:sub_output_dae} is
\begin{align} \label{eq:sub_output_dae_detail}
  \left\{ \begin{aligned}
    \sub{F}_{r}(t, x, \dot{x}, \ldots, , x^{(l + \kappa)}) &= 0,\\
    F_I(t, x, \dot{x}, \ldots, x^{(l)}) &= 0, \\
    F_S(t, x, \dot{x}, \ldots, x^{(l)}) &= 0,
  \end{aligned} \right.
\end{align}
where $\sub{F}_r$ is a function obtained from $F_r$ by substituting~\eqref{eq:phi}.

\subsection[Algorithm for finding (r, I, J)]{Algorithm for Finding $(r, I, J)$}
\label{sec:algorithm_for_finding_rIJ}

Let $D$ be a singular $n \times n$ matrix over a field $\setF$ with row index set $R$ and column index set $C$, and $p = (p_i)_{i \in R}$ an integer vector indexed by $R$.
On the setting in \cref{sec:substitution_method_description}, $\setF$ is the quotient field of the ring of analytic functions on $\setT \times \Omega$.
We give an algorithm, which uses arithmetic operations over $\setF$, to find $r \in R, I \subseteq R \setminus \set{r}$ and $J \subseteq C$ satisfying the conditions (C1)--(C3).

First, by column operations, we transform $D$ into $D' = (D'_{i,j})_{i \in R, j \in C}$ in the form
\begin{align} \label{eq:modified_D}
  D' = \begin{blockarray}{ccc}
    & B & C \setminus B \\
    \begin{block}{c(cc)}
      H             & I & O \\
      R \setminus H & * & O \\
    \end{block}
  \end{blockarray}\,,
\end{align}
where $H \subseteq R$ and $B \subseteq C$ with $\card{H} = \card{B} = \rank D$.
Here, $I$ and $O$ in~\eqref{eq:modified_D} are the identity matrix and the zero matrix of appropriate size, respectively, and ``$*$'' indicates an arbitrary matrix.
Let $\funcdoms{h}{B}{H}$ denote the natural bijection represented by the top left block $D'[H, B]$ in~\eqref{eq:modified_D}.
Namely, $h(j) = i$ if and only if $D'_{i, j} \neq 0$ for $j \in B$ and $i \in H$.

Next, we choose $\ell \in R \setminus H$ arbitrarily.
Note that $R \setminus H$ is nonempty because $D'$ is singular.
Put 
\begin{align} \label{def:Z}
  Z \defeq \set{\ell} \cup \set{h(j)}[j \in B, D'_{\ell,j} \neq 0] \subseteq R.
\end{align}
Finally, we take $r \in Z$ such that $p_r \leq p_i$ for all $i \in Z$.
Put $I \defeq Z \setminus \set{r}$ and choose $J \subseteq C$ such that $D[I, J]$ is nonsingular.
The existence of $J$ is guaranteed by the following lemma.

\begin{lemma} \label{lem:rIJ}
  Let $D \in \setF^{n \times n}$ be a singular matrix and $Z \subseteq R$ defined in~\eqref{def:Z}.
  Then $D[Z, C]$ is not of full-row rank and $D[I, C]$ is of full-row rank for any proper subset $I \subsetneq Z$.
\end{lemma}
\begin{proof}
  Since $D'$ in~\eqref{eq:modified_D} is a matrix obtained from $D$ by column operations, it suffices to show the statement for $D'$.
  By the definition of $Z$, it holds
  \begin{align}
    D'[\set{\ell}, C] - \sum_{i \in Z \setminus \set{\ell}} D'[\set{i}, C] = 0.
  \end{align}
  This implies that $D'[Z, C]$ is not of full-row rank.
  
  We next show that $D'[I, C]$ is of full-row rank for $I \subsetneq Z$.
  This is trivial if $\ell \notin I$ since $I \subsetneq Z \subseteq \set{\ell} \cup H$ and $D'[H, C]$ is of full-row rank.
  Suppose that $\ell \in I$.
  Then we can take $i \in Z \setminus I$.
  From the definition of $Z$, $D'[(Z \setminus \set{i}) \cup \set{\ell}, C]$ is of full-row rank.
  Since $I \subseteq (Z \setminus \set{i}) \cup \set{\ell}$, $D'[I, C]$ is also of full-row rank.
  \qed
\end{proof}

The following theorem holds from the construction of $(r, I, J)$ together with \cref{lem:rIJ}.

\begin{theorem}
  For a singular matrix $D \in \setF^{n \times n}$, the above algorithm returns $(r, I, J)$ satisfying the conditions \textup{(C1)--(C3)}.
\end{theorem}

This algorithm uses $\Order{n^3}$ arithmetic operations over $\setF$.

\subsection{Application of Implicit Function Theorem}
\label{sec:application_of_implicit_function_theorem}

This section gives a mathematically rigorous description of the application of the IFT to~\eqref{eq:F_I_diff}.
The description in this section is used in proofs of the substitution method later.

We introduce additional notations.
Let $\mathcal{C} \subseteq C \times \set{0, 1, 2, \ldots, l}$ be a finite set of index pairs such that $(j, k) \in \mathcal{C}$ indicates an argument $x_j^{(k)}$ of $F$ in~\eqref{eq:dae}.
Let $\setR^{\mathcal{C}}$ denote a $\card{\mathcal{C}}$-dimensional real vector space with index set $\mathcal{C}$.
For $X \in \setR^{\mathcal{C}}$ and $\mathcal{J} \subseteq \mathcal{C}$, let $X_{\mathcal{J}}$ designate a subvector of $X$ with index set $\mathcal{J}$.

The following is a version of the IFT which we use.

\begin{theorem}[{Implicit Function Theorem; IFT}]
  Let $U \subseteq \setR^{n+m}$ be an open set having coordinates $(x, y)$ with $x \in \setR^n$ and $y \in \setR^m$.
  Let $\funcdoms{f}{U}{\setR^m}$ be a real analytic function.
  Fix a point $(\xi, \eta) \in U$ such that $f(\xi, \eta) = 0$ and $\pdif{f}{y}(\xi, \eta)$ is nonsingular.
  Then there exist open sets $V \subseteq \setR^n$ and $W \subseteq \setR^m$ with $(\xi, \eta) \in V \times W \subseteq U$ and a real analytic function $\funcdoms{\phi}{V}{W}$ such that
  \begin{enumerate}[label={\textup{(\arabic*)}}]
    \item $\phi(\xi) = \eta$,
    \item $f(x, y) = 0$ if and only if $y = \phi(x)$ for all $(x, y) \in V \times W$, and
    \item $\pdif{f}{y}(x, \phi(x))$ is nonsingular and
          \begin{align} \label{eq:implicit_differentiation}
            \dif{\phi}{x}(x) = -\prn{\pdif{f}{y}(x, \phi(x))}^{-1} \pdif{f}{x}(x, \phi(x))
          \end{align}
          for all $x \in V$.
  \end{enumerate}
\end{theorem}

The function $\phi$ in the IFT is called an \emph{explicit function}.
The formula~\eqref{eq:implicit_differentiation} is called the \emph{implicit differentiation formula}.

Let us start the description of the application of the implicit function theorem.
Let $(p, q)$ be an optimal solution of~$\D{F}$ and $(r, I, J)$ triple satisfying the conditions (C1)--(C3).
Put
\begin{align} \label{def:mathcal_C}
  \mathcal{C} \defeq \set{(j, k)}[j \in C,\, 0 \leq k \leq q_j - p_r].
\end{align}
From \cref{lem:griewank} and the feasibility of $(p, q)$, it holds that
\begin{align} \label{eq:order_of_F_i_diff}
  \ord{F_i^{(p_i - p_r)}}{x_j} = \ord{F_i}{x_j} + p_i - p_r = c_{i,j} + p_i - p_r \leq q_j - p_r
\end{align}
for $i \in I \cup \set{r}$ and $j \in J$ with $\ord{F_i}{x_j} > -\infty$.
Thus we regard both $F_r$ and $F_I^{(p_I - p_r\onevec)}$ as functions defined on $\setT \times U$, where $U$ is an open subset of $\setR^{\mathcal{C}}$.

Take $(\hat{t}, \hat{X}) \in \setT \times U$ such that $F_I^{(p - p_r\onevec)} (\hat{t}, \hat{X}) = 0$ and $\pdif{F_I^{(p - p_r \onevec)}}{x_J^{(q - p_r \onevec)}}(\hat{t}, \hat{X})$ is nonsingular.
Let
\begin{align} \label{def:mathcal_J}
  \mathcal{J} \defeq \set{(j, q_j - p_r)}[j \in J] \subseteq \mathcal{C}.
\end{align}
Then the components of $\hat{X}$ is bipartitioned by $\mathcal{J}$ as $\hat{X} = (\hat{X}_{\mathcal{C} \setminus \mathcal{J}}, \hat{X}_{\mathcal{J}})$.
Thus by the implicit function theorem, there exist open sets $\sub{\setT} \subseteq \setT$, $V \subseteq \setR^{\mathcal{C} \setminus \mathcal{J}}$ and $W \subseteq \setR^{\mathcal{J}}$ with $(\hat{t}, \hat{X}_{\mathcal{C} \setminus \mathcal{J}}, \hat{X}_{\mathcal{J}}) \in \sub{\setT} \times V \times W \subseteq \setT \times U$ and a real analytic function $\funcdoms{\phi}{\sub{\setT} \times V}{W}$ such that $\hat{X}_{\mathcal{J}} = \app[\big]{\phi}{\hat{t}, \hat{X}_{\mathcal{C} \setminus \mathcal{J}}}$ and
\begin{align} \label{eq:F_I_substituted_phi}
  \app[\big]{F_I^{(p - p_r\onevec)}}{
    t,
    X_{\mathcal{C} \setminus \mathcal{J}},
    \app[\big]{\phi}{t, X_{\mathcal{C} \setminus \mathcal{J}}}
  }
  = 0
\end{align}
for every $(t, X_{\mathcal{C} \setminus \mathcal{J}}) \in V$.
In addition, all zeros of $F_I^{(p - p_r\onevec)}$ in $\sub{\setT} \times V \times W$ are in the form of~\eqref{eq:F_I_substituted_phi}.
Using $\phi$, the modified function $\funcdoms{\sub{F}_r}{\sub{\setT} \times V}{\setR}$ can be expressed as
\begin{align} \label{eq:Fr_sub_by_Fr}
  \sub{F}_r(t, X_{\mathcal{\mathcal{C}} \setminus \mathcal{J}})
  =
  \app[\big]{F_r}{
    t,
    X_{\mathcal{\mathcal{C}} \setminus \mathcal{J}},
    \app[\big]{\phi}{t, X_{\mathcal{C} \setminus \mathcal{J}}}
  }
\end{align}
for $(t, X_{\mathcal{C} \setminus \mathcal{J}}) \in \sub{\setT} \times V$.
Since both $F_r$ and $\phi$ are real analytic, so is $\sub{F}_r$.

We remark about the domain of the resulting system of functions $\bar{F}$ in~\eqref{eq:sub_output_dae_detail}.
In the above argument, we treated the domain of $F_I^{(p - p_r\onevec)}$ as $\setT \times U$, which is an open subset of $\setT \times \setR^{\mathcal{C}}$.
However, the domain of $F_I^{(p - p_r\onevec)}$ can also be represented as $\setT \times \Omega^{(\kappa)}$, where $\kappa$ is defined by~\eqref{def:kappa} (indeed, $U$ is the projection of $\Omega^{(\kappa)}$ onto $\setR^{\mathcal{C}}$).
Since $\bar{F}_r$ is a function obtained from $F_I^{(p - p_r\onevec)}$ and $F_r$ by the above transformation, the domain of $\bar{F}_r$ (and thus of $\bar{F}$) can also be regarded as $\sub{\setT} \times \sub{\Omega}$, where $\sub{\Omega}$ is a nonempty open subset of $\Omega^{(\kappa)}$.

\subsection{Proofs}

This section is devoted to the validity proofs of our method.

We first show (S1).
In \cref{sec:application_of_implicit_function_theorem}, we have already shown that $F_r$ is a real analytic function defined on $\sub{\setT} \times \sub{\Omega}$.
Thus what we should give is only the bound on $\kappa$.
To give this bound, we need to bound $p_i$ for $i \in I$.
This can be achieved by using a specialized $(p, q)$ that the following lemma claims.

\begin{lemma}[{\cite[Lemma~4.1]{Iwata2018b}}] \label{lem:bound_on_pq}
  Let $G = (R \cup C; E)$ be a bipartite graph with $\card{R} = \card{C} = n$ and suppose that $G$ has a perfect matching.
  For each edge $e \in E$, let $c_e \in \set{0, 1, \ldots, l}$ be the weight of $e$.
  Then there exists an algorithm to find an optimal solution $(p, q)$ of the dual of the maximum weight perfect matching problem on $G$ such that $p_i, q_j \leq nl$ for all $i \in R$ and $j \in C$.
\end{lemma}

See \cite[Section~4.2]{Iwata2018b} for the algorithm mentioned in \cref{lem:bound_on_pq}.
By using $(p, q)$ obtained by the algorithm, the following lemma immediately follows.

\begin{lemma} \label{lem:kappa_bound}
  In the substitution method, $\kappa$ defined in~\eqref{def:kappa} is at most $nl$.
\end{lemma}

Next, we focus on (S2), which claims about the equivalence of the original DAE and the modified DAE.

\begin{lemma} \label{lem:S2}
  Consider a DAE~\eqref{eq:dae} satisfying \textup{(I1)--(I3)}.
  Let $\funcdoms{x}{\sub{\setT}}{\setR^n}$ be a sufficiently smooth trajectory satisfying the initial value condition~\eqref{eq:initial_value_condition} for $(t^*, X^*)\in \sub{\setT} \times \sub{\Omega}$.
  Then there exists an open subinterval $\setI \subseteq \sub{\setT}$ containing $t^*$ such that $x$ is a solution of~\eqref{eq:dae} on $\setI$ if and only if $x$ is a solution of~\eqref{eq:sub_output_dae_detail} on $\setI$.
\end{lemma}
\begin{proof}
  We show both the ``if'' and ``only if'' parts simultaneously.
  Suppose that there exists an open subinterval $\setI \subseteq \sub{\setT}$ with $t^* \in \setI$ such that $x$ is a solution of~\eqref{eq:dae} or~\eqref{eq:sub_output_dae_detail} on $\setI$.
  Then $x$ satisfies $F_I(t, x(t), \dot{x}(t), \ldots, x^{(l)}(t)) = 0$ on $\setI$, which is a subsystem of both~\eqref{eq:fHK} and~\eqref{eq:sub_output_dae_detail}.
  Thus $x$ also satisfies~\eqref{eq:F_I_diff} on $\setI$.
  
  We rewrite the equation~\eqref{eq:F_I_diff} for $x(t)$ using $\mathcal{C}$ and $\mathcal{J}$ defined by~\eqref{def:mathcal_C} and~\eqref{def:mathcal_J}, respectively.
  Let $\difop{\mathcal{C}}$ denote a differentiation operator that maps $x$ to a trajectory $\funcdoms{\difop{\mathcal{C}}x}{\sub{\setT}}{\setR^{\mathcal{C}}}$ such that
  \begin{align} \label{def:difop}
    \prn{\difop{\mathcal{C}}x(t)}_{(j, k)} = x_j^{(k)}(t)
  \end{align}
  for $t \in \sub{\setT}$ and $(j, k) \in \mathcal{C}$.
  Then the initial value condition~\eqref{eq:initial_value_condition} can be represented as $\difop{\mathcal{C}} x(t^*) = X^*_{\mathcal{C}}$.
  Since the domain of $F_I^{(p - p_r\onevec)}$ is an open subset of $\setT \times \setR^{\mathcal{C}}$, the equation~\eqref{eq:F_I_diff} for $x(t)$ can also be represented as
  \begin{align}
    \app[\big]{F_I^{(p - p_r\onevec)}}{
      t,
      \difop{\mathcal{C}}x(t)
    }
    = 0
  \end{align}
  or
  \begin{align} \label{eq:F_I_D_x}
    \app[\big]{F_I^{(p - p_r\onevec)}}{
      t,
      \difop{\mathcal{C} \setminus \mathcal{J}}x(t),
      \difop{\mathcal{J}}x(t)
    }
    = 0
  \end{align}
  for $t \in \setI$, where $\difop{\mathcal{C} \setminus \mathcal{J}}$ and $\difop{\mathcal{J}}$ are differentiation operators defined in the same way as $\difop{\mathcal{C}}$.
  
  Let $U \subseteq \setR^{\mathcal{C}}, V \subseteq \setR^{\mathcal{C} \setminus \mathcal{J}}$ and $W \subseteq \setR^{\mathcal{J}}$ be open sets defined in \cref{sec:application_of_implicit_function_theorem}.
  Here, since $x$ is smooth, $U$ is open and $\difop{\mathcal{C}} x(t^*) = X^*_{\mathcal{C}} \in U$, it holds $\difop{\mathcal{C}} x(t) \in U$ for all $t \in \setI$ by taking $\setI$ sufficiently small.
  This implies that $\difop{\mathcal{C} \setminus \mathcal{J}}x(t) \in V$ and $\difop{\mathcal{J}}x(t) \in W$ for $t \in \setI$.
  Comparing~\eqref{eq:F_I_substituted_phi} and~\eqref{eq:F_I_D_x}, we obtain
  \begin{align} \label{eq:x_implicit_function}
    \difop{\mathcal{J}}x(t) = \app[\big]{\phi}{t, \difop{\mathcal{C} \setminus \mathcal{J}}x(t)}
  \end{align}
  for $t \in \setI$.
  Therefore, we have
  \begin{align}
    F_r(t, x(t), \dot{x}(t), \ldots, x^{(l)}(t))
    &=
    \app[\big]{F_r}{
      t,
      \difop{\mathcal{C} \setminus \mathcal{J}} x(t),
      \difop{\mathcal{J}}x(t)
    } \\
    &= 
    \app[\big]{F_r}{
      t,
      \difop{\mathcal{C} \setminus \mathcal{J}} x(t),
      \app[\big]{\phi}{t, \difop{\mathcal{C} \setminus \mathcal{J}} x(t)}
    } \\
    &=
    \app[\big]{\sub{F}_r}{t, \difop{\mathcal{C} \setminus \mathcal{J}}x(t)} \\
    &=
    \sub{F}_r(t, x(t), \dot{x}(t), \ldots, x^{(l + \kappa)}(t)),
  \end{align}
  which means that $x$ is a solution of~\eqref{eq:sub_output_dae_detail} if $x$ is a solution of~\eqref{eq:dae}, and vice versa.
  \qed
\end{proof}

We finally show that the modified DAE satisfies (S3).
In order to show (S3), it suffices to show that $(p, q)$ is a feasible solution of $\D{\sub{F}}$ but not an optimal solution.
The feasibility is easily shown as follows.

\begin{lemma} \label{lem:modified_pq_feasible}
  Consider a DAE~\eqref{eq:dae} satisfying \textup{(I1)--(I3)} and let $(p, q)$ be an optimal solution of $\D{F}$.
  Then $(p, q)$ is feasible on $\D{\sub{F}}$.
\end{lemma}
\begin{proof}
  Let $(r, I, J)$ be a triple satisfying the conditions (C1)--(C3).
  Consider the explicit function $\phi$ in~\eqref{eq:phi}.
  For $i \in I$ and $j \in C$, we have
  \begin{align}
    \ord{\phi}{x_j}
    \leq
    \ord{F_i^{(p_i - p_r)}}{x_j}
    \leq
    q_j - p_r,
  \end{align}
  where we used~\eqref{eq:order_of_F_i_diff} in the last inequality.
  Because $\sub{F}_r$ is a function obtained from $F_r$ by substituting~\eqref{eq:phi}, it holds
  \begin{align}
    \ord{\sub{F}_r}{x_j} \leq \max \set{\ord{F_r}{x_j}, \ord{\phi}{x_j}} \leq q_j - p_r
  \end{align}
  for every $j \in C$.
  Thus $(p, q)$ is feasible on $\D{\sub{F}}$.
  \qed
\end{proof}

We finally focus on the non-optimality of $(p, q)$ on $\D{\sub{F}}$.
By \cref{prop:optimality}, our goal is to show that $\trank \bar{D} < n$ holds, where $\bar{D}$ be the system Jacobian of~\eqref{eq:sub_output_dae} with respect to $(p, q)$.
This is shown by the following lemma.

\begin{lemma} \label{lem:tight_variables_vanish}
  Consider a DAE~\eqref{eq:dae} satisfying \textup{(I1)--(I3)}.
  Let $(p, q)$ be an optimal solution of $\D{F}$ and $(r, I, J)$ a triple satisfying \textup{(C1)--(C3)}.
  Then the modified function $\bar{F}_r$ in~\eqref{eq:sub_output_dae_detail} does not depend on $x_j^{(q_j - p_r)}$ for all $j \in C$.
\end{lemma}
\begin{proof}
  The claim is easy to see for $j \in J$ because we have eliminated $x_j^{(q_j - p_r)}$ from $F_r$ by substituting~\eqref{eq:phi}.
  Consider the variable $x_T^{(q - p_r \onevec)}$ with $T = C \setminus J$.
  Let $\mathcal{C}$ and $\mathcal{J}$ be index sets defined in~\eqref{def:mathcal_C} and~\eqref{def:mathcal_J}, respectively.
  For $(t, X) \in \sub{\setT} \times \sub{\Omega}$, we denote $(t, X_{\mathcal{C} \setminus \mathcal{J}}, \phi(t, X_{\mathcal{C} \setminus \mathcal{J}}))$ by $A_{t, X}$ for short, where $\phi$ is the explicit function given by~\eqref{eq:phi}.
  From the chain rule, the implicit differentiation formula~\eqref{eq:implicit_differentiation} and \Cref{lem:griewank}, we obtain
  \begin{align}
    &\pdif{\sub{F}_r}{x^{(q - p_r\onevec)}}(t, X_{\mathcal{C} \setminus \mathcal{J}}) \\
    &=
    \pdif{F_r}{x_{T}^{(q - p_r\onevec)}}(A_{t, X}) + \pdif{F_r}{x_J^{(q - p_r \onevec)}}(A_{t, X}) \pdif{\phi}{x_{T}^{(q - p_r\onevec)}}(t, X_{\mathcal{C} \setminus \mathcal{J}}) \\
    &=
    \pdif{F_r}{x_{T}^{(q - p_r\onevec)}}(A_{t, X}) - \pdif{F_r}{x_J^{(q - p_r \onevec)}}(A_{t, X}) \prn{\pdif{F_I^{(p - p_r\onevec)}}{x_J^{(q - p_r \onevec)}}(A_{t, X})}^{-1} \pdif{F_I^{(p - p_r\onevec)}}{x_{T}^{(q - p_r\onevec)}}(A_{t, X}) \\
    &=
    \pdif{F_r^{(p_r)}}{x_{T}^{(q)}}(A_{t, X}) - \pdif{F_r^{(p_r)}}{x_J^{(q)}}(A_{t, X}) \prn{\pdif{F_I^{(p)}}{x_J^{(q)}}(A_{t, X})}^{-1} \pdif{F_I^{(p)}}{x^{(q)}}(A_{t, X})
    \label{eq:schur}
  \end{align}
  for $(t, X) \in \sub{\mathbb{T}} \times \sub{\Omega}$.
  The right hand side of~\eqref{eq:schur} coincides with the Schur complement of $\pdif{F_I^{(p)}}{x_J^{(q)}}(A_{t, X})$ in the following matrix
  \begin{align}
    \tilde{D}(t, X_{\mathcal{C} \setminus \mathcal{J}})
    \defeq
    \begin{pmatrix}
      \pdif{F_r^{(p_r)}}{x_J^{(q)}}(A_{t, X}) & \pdif{F_r^{(p_r)}}{x_T^{(q)}}(A_{t, X}) \\
      \pdif{F_I^{(p)}}{x_J^{(q)}}(A_{t, X}) & \pdif{F_I^{(p)}}{x_T^{(q)}}(A_{t, X}) \\
    \end{pmatrix}.
  \end{align}
  Thus, we have
  \begin{align} \label{eq:ranks}
    \rank \tilde{D}(t, X_{\mathcal{C} \setminus \mathcal{J}}) = \rank \pdif{F_I^{(p)}}{x_J^{(q)}}(A_{t, X}) + \rank \pdif{\sub{F}_r}{x_{T}^{(q - p_r\onevec)}}(t, X_{\mathcal{C} \setminus \mathcal{J}})
  \end{align}
  for all $(t, X) \in \sub{\mathbb{T}} \times \sub{\Omega}$.
  Let $D$ be a system Jacobian of $F$ with respect to $(p, q)$.
  Note that $\tilde{D}$ is a matrix obtained from $D[I \cup \set{r}, C]$ by substituting $\phi(t, X_{\mathcal{C} \setminus \mathcal{J}})$ into $X_{\mathcal{J}}$.
  Hence it holds $\rank \tilde{D}(t, X_{\mathcal{C} \setminus \mathcal{J}}) \leq \rank D[I \cup \set{r}, C](t, X) \leq \rank D[I \cup \set{r}, C] = m$ with $m = \card{I}$, where the last equality comes from (C2).
  In addition, the rank of $\pdif{F_I^{(p)}}{x_J^{(q)}}(A_{t, X})$ is $m$ due to the invertibility.
  Therefore, the rank of $\pdif{\sub{F}_r}{x_{T}^{(q - p_r\onevec)}}(t, X_{\mathcal{C} \setminus \mathcal{J}})$ is zero from~\eqref{eq:ranks}, which means that $\pdif{\sub{F}_r}{x_{T}^{(q - p_r\onevec)}}$ is identically zero on $\sub{\setT} \times \sub{\Omega}$.
  Thus $\sub{F}_r$ does not depend on $x_j^{(q_j - p_r)}$ for $j \in T$.
  \qed
\end{proof}

\begin{corollary} \label{lem:nonlinear_not_feasible}
  For a DAE~\eqref{eq:dae} satisfying \textup{(I1)--(I3)}, it holds $\tdegdet{\sub{F}} \leq \tdegdet{F} - 1$.
\end{corollary}
\begin{proof}
  Let $(p, q)$ be an optimal solution of $\D{F}$ and $(r, I, J)$ a triple satisfying the conditions (C1)--(C3).
  Let $\bar{D}$ be the system Jacobian of~\eqref{eq:sub_output_dae_detail} with respect to $(p, q)$.
  By \Cref{lem:griewank}, it holds
  \begin{align}
    \bar{D}[\set{r}, C] = \pdif{\bar{F}_r^{(p_r)}}{x^{(q)}} = \pdif{\bar{F}_r}{x^{(q - p_r \onevec)}},
  \end{align}
  whereas the right-hand side is identically zero from~\Cref{lem:tight_variables_vanish}.
  Thus $\trank \bar{D}$ is less than $n$, and from \Cref{prop:optimality}, $(p, q)$ is not an optimal solution of $\D{\sub{F}}$.
  This concludes the proof.
  \qed
\end{proof}

We conclude this section with the following theorem.

\begin{theorem}
  For a DAE~\eqref{eq:dae} satisfying \textup{(I1)--(I3)}, the substitution method outputs a DAE~\eqref{eq:sub_output_dae} satisfying \textup{(S1)--(S3)}.
\end{theorem}

\section{Augmentation Method}
\label{sec:augmentation_method}

\subsection{Method Description}

This section describes another proposed modification method for nonlinear DAEs, which we call an augmentation method.
The input of the augmentation method is a nonlinear DAE~\eqref{eq:dae} of size $n$ satisfying the conditions (I1)--(I3), where $\funcdoms{F}{\setT \times \Omega}{\setR^n}$ is a real analytic function again.
Instead of solving equations symbolically, the augmentation method augments the size of the DAE by introducing a new variable vector $y$ and attaching new equations.
Formally, the augmentation method modifies~\eqref{eq:dae} into a DAE
\begin{align} \label{eq:aug_output_dae}
  \aug{F}(t, x, \dot{x}, \ldots, x^{(l+\kappa)}, y) = 0
\end{align}
of size $n + m$ such that
\begin{enumerate}[label={(A\arabic*)}]
  \item $\aug{F}$ is a real analytic function defined on a nonempty open subset $\aug{\setT} \times \aug{\Omega} \times Y \subseteq \setT \times \Omega^{(\kappa)} \times \setR^m$ with $\kappa \leq ln$ and $m \leq n-1$,
  \item the resulting DAE~\eqref{eq:aug_output_dae} is locally equivalent to~\eqref{eq:dae}, and
  \item $\tdegdet{\aug{F}} \leq \tdegdet{F} - 1$.
\end{enumerate}
See \cref{lem:A2} for the precise meaning of ``locally equivalent'' in (A2).

The substitution method and the augmentation method are the same except for the last modification process.
The overlapping part is described here briefly.
Let $R$ and $C$ be the equation index set and the variable index set of the input DAE~\eqref{eq:dae}, respectively.
Let $(p, q)$ be an optimal solution of $\D{F}$ and $D$ denote the system Jacobian with respect to $(p, q)$.
We first find $r \in R$, $I \subseteq R \setminus \set{r}$ and $J \subseteq C$ satisfying the conditions (C1)--(C3) described in \cref{sec:substitution_method_description}.
Define $\kappa \defeq \displaystyle \max_{i \in I} p_i - p_r, m \defeq \card{I}, S \defeq R \setminus (I \cup \set{r})$ and $T \defeq C \setminus J$.

The following modification step differs from the substitution method.
Let $I' = \set{i'}[i \in I]$ and $J' = \set{j'}[j \in J]$ be copies of $I$ and $J$, respectively.
Take a point $(\tau, \Xi)$ arbitrary from the domain $\sub{\setT} \times \sub{\Omega} \subseteq \setT \times \Omega^{(\kappa)}$ of the resultant DAE $\sub{F}$ of the substitution method.
We regard $\Omega^{(\kappa)}$ as a subset of $\setR^{\mathcal{C}}$ hereafter, where $\mathcal{C} \defeq C \times \set{0, 1, 2, \ldots, l+\kappa}$.
For $X \in \setR^{\mathcal{C}}$ and a vector $y = (y_{j'})_{j' \in J'}$ with index set $J'$, let $\psi_\Xi(X, y)$ be a vector of $\setR^{\mathcal{C}}$ such that
\begin{align}
  \prn{\psi_\Xi(X, y)}_{(j, k)}
  \defeq \begin{cases}
      y_{j'}     & (j \in J, k = q_j - p_r), \\
    \Xi_{(j, k)} & (j \in T, k = q_j - p_r), \\
      X_{(j, k)} & (\text{otherwise})
  \end{cases}
\end{align}
for $(j, k) \in \mathcal{C}$.
For each $i \in I$, we define a function
\begin{align}
  \aug{F}_{i'}(t, x, \dot{x}, \ldots, x^{(l+\kappa)}, y)
  \defeq
  F_i^{(p_i - p_r)}(t, \psi_\Xi(X, y)),
\end{align}
where $X = (x, \dot{x}, \ldots, x^{(l+\kappa)})$.
Namely, $\aug{F}_{i'}$ is obtained by replacing $x_j^{(q_j-p_r)}$ in $F_i^{(p_i - p_r)}$ with a variable $y_{j'}$ for $j \in J$ and with a constant $\Xi_{(j, q_j-p_r)}$ for $j \in T$.
Put $\aug{F}_{I'} \defeq \prn{\aug{F}_{i'}}_{i' \in I'}$.
We also define
\begin{align}
  \aug{F}_r(t, x, \dot{x}, \ldots, x^{(l+\kappa)}, y)
  \defeq
  F_r(t, \psi_\Xi(X, y))
\end{align}
in the same way.

The output~\eqref{eq:aug_output_dae} of the augmentation method is the following DAE
\begin{align} \label{eq:aug_output_dae_detail}
  \left\{ \begin{aligned}
    \aug{F}_{r}(t, x, \dot{x}, \ldots, , x^{(l + \kappa)}, y) &= 0,\\
    F_I(t, x, \dot{x}, \ldots, x^{(l)}) &= 0, \\
    F_S(t, x, \dot{x}, \ldots, x^{(l)}) &= 0, \\
    \aug{F}_{I'}(t, x, \dot{x}, \ldots, x^{(l + \kappa)}, y) &= 0
  \end{aligned} \right.
\end{align}
with unknown function $(x(t), y(t))$ of $t$.
The domain $\aug{\setT} \times \aug{\Omega}$ of~\eqref{eq:aug_output_dae_detail} is given by $\aug{\setT} \defeq \sub{\setT}$ and $\aug{\Omega} \defeq \set{(X, y) \in \sub{\Omega} \times \setR^{m}}[\psi_\Xi(X, y) \in \sub{\Omega}]$.

The DAE~\eqref{eq:aug_output_dae_detail} is obtained by copying some equations (or their derivatives), relabelling variables and substituting constants.
Hence if the original DAE contains only a few variables in each equation, so does~\eqref{eq:aug_output_dae_detail}.
Thus the augmentation method retains the sparsity of DAEs.

\subsection{Proofs}

Validity proofs of the augmentation method are given in this section. 
We first show (A1).

\begin{lemma}
  For a DAE~\eqref{eq:dae} satisfying \textup{(I1)--(I3)}, the resulting DAE $\aug{F} = 0$ satisfies \textup{(A1)}.
\end{lemma}
\begin{proof}
  It is clear that $\aug{F}$ is real analytic from its construction, which is a combination of variable relabelling and partial substitution of constants on $F$.
  Let $\eta = (\eta_{j'})_{j' \in J'}$ be a vector defined by $\eta_{j'} = \Xi_{(j, q_j-p_r)}$ for $j' \in J'$.
  Then it holds $(\Xi, \eta) \in \aug{\Omega}$ from $\psi_\Xi(\Xi, \eta) = \Xi \in \sub{\Omega}$.
  Hence $\aug{\Omega}$ is nonempty.
  In addition, since $\psi_\Xi$ is a continuous map and $\sub{\Omega}$ is an open set, $\aug{\Omega}$ is also open.
  Therefore the domain $\aug{\mathbb{T}} \times \aug{\Omega}$ of $\aug{F}$ is a nonempty open set.
  
  The bounds on $\kappa$ and $m$ are given by \cref{lem:kappa_bound} and $m = \card{I} \leq n - 1$.
  \qed
\end{proof}

We next show (A2) in the sense of the following lemma.

\begin{lemma} \label{lem:A2}
  Consider a DAE~\eqref{eq:dae} satisfying \textup{(I1)--(I3)}.
  Let $\funcdoms{x}{\aug{\setT}}{\setR^n}$ be a sufficiently smooth trajectory satisfying the initial value condition~\eqref{eq:initial_value_condition} for $(t^*, X^*) \in \aug{\setT} \times \aug{\Omega}$.
  Then there exists an open subinterval $\setI \subseteq \aug{\setT}$ containing $t^*$ such that the following two statements are equivalent:
  \begin{enumerate}[label={\textup{(\arabic*)}}]
    \item $x$ is a solution of~\eqref{eq:dae} on $\setI$, and
    \item there uniquely exists a trajectory $\funcdoms{y}{\setI}{\setR^m}$ such that $(x, y)$ is a solution of~\eqref{eq:aug_output_dae_detail} on $\setI$.
  \end{enumerate}
\end{lemma}
\begin{proof}
  From the argument on the substitution method, the last equation $\aug{F}_{I'}(t, x, \dot{x}, \ldots, x^{(l + \kappa)}, y) = 0$ in~\eqref{eq:aug_output_dae_detail} can be solved for $y$ on the domain of $\aug{F}$ as
  \begin{align}
    y = \aug{\phi}(t, x, \dot{x}, \ldots, x^{(l+\kappa)}),
  \end{align}
  where $\aug{\phi}$ is a function obtained by replacing $x_j^{(q_j - p_r)}$ of $\phi$ in~\eqref{eq:phi} with the constant $\Xi_{(j, q_j-p_r)}$ for $j \in T$.
  Therefore,~\eqref{eq:aug_output_dae_detail} is equivalent to
  \begin{align} \label{eq:A2_proof_1}
    \left\{ \begin{aligned}
      \aug{F}_{r}(t, x, \dot{x}, \ldots, , x^{(l + \kappa)}, \aug{\phi}(t, x, \dot{x}, \ldots, x^{(l+\kappa)})) &= 0,\\
      F_I(t, x, \dot{x}, \ldots, x^{(l)}) &= 0, \\
      F_S(t, x, \dot{x}, \ldots, x^{(l)}) &= 0, \\
      y &= \aug{\phi}(t, x, \dot{x}, \ldots, x^{(l+\kappa)}).
    \end{aligned} \right.
  \end{align}
  It can be seen from~\eqref{eq:Fr_sub_by_Fr} that the left-hand side of the first equation in~\eqref{eq:A2_proof_1} is a function obtained by replacing $x_j^{(q_j - p_r)}$ of $\sub{F}_r$ with the constant $\Xi_{(j, q_j-p_r)}$ for $j \in T$.
  On the other hand, $\sub{F}_r$ does not depend on $x_j^{(q_j - p_r)}$ for all $j \in T$ from \cref{lem:tight_variables_vanish}.
  Therefore, the first equation in~\eqref{eq:A2_proof_1} is equivalent to $\sub{F}_r(t, x, \dot{x}, \ldots, , x^{(l + \kappa)}) = 0$.
  Thus the system~\eqref{eq:A2_proof_1} is equivalent to
  \begin{align} \label{eq:A2_proof_2}
    \left\{ \begin{aligned}
      \sub{F}(t, x, \dot{x}, \ldots, , x^{(l + \kappa)}) &= 0, \\
      y &= \aug{\phi}(t, x, \dot{x}, \ldots, x^{(l+\kappa)}).
    \end{aligned} \right.
  \end{align}
  
  The statement of this lemma is shown by~\eqref{eq:A2_proof_2} together with \cref{lem:S2}.
  \qed
\end{proof}

Let $\bar{R} \defeq R \cup I'$ and $\bar{C} \defeq C \cup J'$.
We finally show (A3) as a corollary of the following lemma.

\begin{lemma} \label{lem:aug_pq}
  Consider a DAE~\eqref{eq:dae} satisfying \textup{(I1)--(I3)} and let $(p, q)$ be a dual optimal solution.
  Define
  \begin{align} \label{eq:aug_pq}
    \bar{p}_i \defeq \begin{cases}
      p_i & (i \in R), \\
      p_r & (i \in I'),
    \end{cases}
    \quad
    \bar{q}_j \defeq \begin{cases}
      q_j & (j \in C), \\
      p_r & (j \in J')
    \end{cases}
  \end{align}
  for $i \in \bar{R}$ and $j \in \bar{C}$.
  Then $(\bar{p}, \bar{q})$ is feasible but not optimal on $\D{\aug{F}}$.
\end{lemma}
\begin{proof}
  We first prove
  \begin{align} \label{eq:lem_feasibility_eq}
    \ord{\aug{F}_{i'}}{x_j} < q_j - p_r
  \end{align}
  for $i' \in I' \cup \set{r}$ and $j \in C$.
  Since $x_j^{(q_j - p_r)}$ in $\aug{F}_{i'}$ has been replaced with a dummy variable or a constant, it holds $\ord{\aug{F}_{i'}}{x_j} < \ord{F_i^{(p_i - p_r)}}{x_j}$, where $i = r$ if $i' = r$ here.
  In addition, $\ord{F_i^{(p_i - p_r)}}{x_j} = \ord{F_i}{x_j} + p_i - p_r \leq q_j - p_r$ holds, where the first equality comes from \cref{lem:griewank} and the second inequality is due to the feasibility of $(p, q)$ on $\D{F}$.
  Thus~\eqref{eq:lem_feasibility_eq} is true.

  We next show the feasibility of $(\bar{p}, \bar{q})$ on $\D{\aug{F}}$.
  For $i \in R \setminus \set{r}$ and $j \in C$, it holds $\ord{\aug{F}_i}{x_j} = \ord{F_i}{x_j} \leq q_j - p_i = \bar{q}_j - \bar{p}_i$ from the feasibility of $(p, q)$ on $\D{F}$.
  For $i \in R \setminus \set{r}$ and $j' \in J'$, we have $\ord{\aug{F}_i}{y_{j'}} = \ord{F_i}{y_{j'}} = -\infty \leq \bar{q}_j - \bar{p}_i$.
  For $i' \in I' \cup \set{r}$ and $j \in C$, it holds $\ord{\aug{F}_{i'}}{x_j} < q_j - p_r = \bar{q}_j - \bar{p}_{i'}$ from~\eqref{eq:lem_feasibility_eq}.
  In the last case with $i' \in I' \cup \set{r}$ and $j' \in J'$, we have $\ord{\aug{F}_{i'}}{y_{j'}} = 0 = p_r - p_r = \bar{p}_{i'} - \bar{q}_{j'}$.
  Thus $(\bar{p}, \bar{q})$ is feasible on $\D{\aug{F}}$.
  
  Finally, we show the non-optimality of $(\bar{p}, \bar{q})$ on $\D{\aug{F}}$.
  From \cref{prop:optimality}, it suffices to show $\trank \bar{D} < n+m$, where $\bar{D}$ is the system Jacobian of~\eqref{eq:aug_output_dae_detail} with respect to $(\bar{p}, \bar{q})$.
  Here, $\bar{D}_{i', j}$ is identically zero for $i' \in I' \cup \set{r}$ and $j \in C$ due to~\eqref{eq:lem_feasibility_eq}.
  \cref{fig:D_bar} shows the zero/nonzero pattern of $\bar{D}$, where $\bar{D}[I, J'] = O$ and $\bar{D}[S, J'] = O$ can also be checked from the definition of $\aug{F}$. 
  Therefore, $I \cup S \cup J'$ is a vertex cover in the corresponding bipartite graph $G^*(D) = (\bar{R} \cup \bar{C}, E^*(D))$ with edge set~\eqref{def:edges}.
  By the K\"{o}nig--Egev\'{a}ry theorem, we have
  \begin{align}
    \trank \bar{D}
    \leq \card{I \cup S \cup J'}
    = m + (n-m-1)+m
    = n+m-1,
  \end{align}
  which completes the proof.
  \qed
\end{proof}

\begin{figure}[tbp]
  \centering
  \begin{tikzpicture}[scale=0.6]
    \draw[very thick] (0,0) -- (8,0) -- (8,8) -- (0,8) -- (0,0);
    \draw (0, 2) -- (8, 2);
    \draw (0, 5) -- (8, 5);
    \draw (0, 7) -- (8, 7);
    \draw (2, 0) -- (2, 8);
    \draw (6, 0) -- (6, 8);
    \node at (-0.5, 7.5) {$r$};
    \node at (-0.5, 6)   {$I$};
    \node at (-0.5, 3.5) {$S$};
    \node at (-0.5, 1)   {$I'$};
    \node at (1   , 8.5) {$J$};
    \node at (4   , 8.5) {$T$};
    \node at (7   , 8.5) {$J'$};
    \node at (1   , 7.5) {$O$};
    \node at (4   , 7.5) {$O$};
    \node at (1   , 1  ) {$O$};
    \node at (4   , 1  ) {$O$};
    \node at (7   , 6  ) {$O$};
    \node at (7   , 3.5) {$O$};
    \fill[pattern=crosshatch dots] (0,7) -- (6,7) -- (6,2) -- (0,2);
    \fill[pattern=crosshatch dots] (6,7) -- (6,8) -- (8,8) -- (8,7);
    \fill[pattern=crosshatch dots] (6,2) -- (8,2) -- (8,0) -- (6,0);
  \end{tikzpicture}
  \caption{%
    The zero/nonzero pattern of the system Jacobian $\bar{D}$ of $\aug{F}$.
    The hatched region may contain nonzero elements.
  }
  \label{fig:D_bar}
\end{figure}

\begin{corollary}
  For a DAE~\eqref{eq:dae} satisfying \textup{(I1)--(I3)}, the resulting DAE $\aug{F} = 0$ satisfies (A3).
\end{corollary}
\begin{proof}
  Let $(p, q)$ be a dual optimal solution on $\D{F}$ and $(\bar{p}, \bar{q})$ defined by~\eqref{eq:aug_pq}.
  From \cref{lem:aug_pq}, it holds that
  \begin{align}
    \tdegdet{\aug{F}}
    &< \sum_{j \in \bar{C}} \bar{q}_j - \sum_{i \in \bar{R}} \bar{p}_i 
    = \prn{m p_r + \sum_{j \in C} q_j} - \prn{m p_r + \sum_{i \in R} p_i}
    = \sum_{j \in C} q_j - \sum_{i \in R} p_i \\
    &= \tdegdet{F}
  \end{align}
  as required.
  \qed
\end{proof}

The above lemmas are summed up in the following theorem.

\begin{theorem}
  For a DAE~\eqref{eq:dae} satisfying \textup{(I1)--(I3)}, the augmentation method returns a DAE~\eqref{eq:aug_output_dae} satisfying \textup{(A1)--(A3)}.
\end{theorem}

\section{Examples}
\label{sec:examples}

In this section, we demonstrate our methods using two examples.

\subsection{Failure Example for LC-method}
\label{sec:failure_example_for_LC_method}

We first apply our modification methods to the index-1 DAE~\eqref{dae:LC_failure}.
Neither the LC-method nor the ES-method works for~\eqref{dae:LC_failure}.

In Phase~1 of the combinatorial relaxation, we find a dual optimal solution $p = (0, 0)$ and $q = (1, 1)$.
The system Jacobian with respect to $(p, q)$ is
\begin{align}
  D = \begin{pmatrix}
    \dot{x}_2 & \dot{x}_1 \\
    2 \dot{x}_1 {{}\dot{x}_2}^2 & 2 {{}\dot{x}_1}^2 \dot{x}_2
  \end{pmatrix},
\end{align}
which is identically singular.
In Phase~3, we find $r = 2, I = \set{1}$ and $J = \set{1}$ as follows:
\begin{align}
  D = \begin{blockarray}{ccc}
    \overmat{1}{2x_1 \dot{x}_2^2}{J} \\
    \begin{block}{(cc)c}
      \dot{x}_2            & \dot{x}_1      & I \\
      2 \dot{x}_1 {{}\dot{x}_2}^2 & 2 {{}\dot{x}_1}^2 \dot{x}_2& r \\
    \end{block}
  \end{blockarray}
  .
\end{align}

We first demonstrate the substitution method applied to~\eqref{dae:LC_failure}.
By solving $F_1 = 0$ in~\eqref{dae:LC_failure} for $\dot{x}_1$, we get
\begin{align} \label{eq:example1_solved}
  \dot{x}_1 = - \frac{2 \cos^2 t}{\dot{x}_2}
\end{align}
unless $\dot{x}_2 = 0$.
Then~\eqref{eq:example1_solved} is substituted into the second equation in~\eqref{dae:LC_failure}.
The resulting DAE of the substitution method is
\begin{align} \label{eq:example1_dae_result}
  \left\{\begin{aligned}
    F_1^{\phantom{\mathrm{sub}}}:      && \dot{x}_1 \dot{x}_2 -2 \cos^2 t &= 0, \\
    \sub{F}_2:&& x_1 + x_2 - 3 \sin t - 2 &= 0.
  \end{aligned}\right.
\end{align}
The system Jacobian $D'$ of~\eqref{eq:example1_dae_result} corresponding to a dual optimal solution $p' = (0, 1), q' = (1, 1)$ is
\begin{align}
  D' = \begin{pmatrix}
    \dot{x}_2 & \dot{x}_1 \\
    1 & 1
  \end{pmatrix},
\end{align}
which is not identically singular.
Thus we are done.

Next, we show the modification of~\eqref{dae:LC_failure} by the augmentation method.
Let $y_{1'}$ be a new variable corresponding to $\dot{x}_1$ and $\xi \in \setR$ an arbitrary nonzero constant corresponding to $\dot{x}_2$.
The augmentation method modifies the DAE~\eqref{dae:LC_failure} into the following DAE
\begin{align}
  \left\{\begin{aligned}
    F_1^{\phantom{\mathrm{aug}}}:&& \dot{x}_1 \dot{x}_2 -2 \cos^2 t &= 0, \\
    \aug{F}_2:&& y_{1'}^2 \xi^2 + x_1 + x_2 - 4 \cos^4 t - 3 \sin t - 2 &= 0, \\
    \aug{F}_{1'}:&& y_{1'} \xi -2 \cos^2 t &= 0 \\
  \end{aligned}\right.
\end{align}
with unknown function $(x_1, x_2, y_{1'})$.
A pair of $p' = (0,1,1)$ and $q' = (1,1,1)$ is a new dual optimal solution, and the system Jacobian $D'$ is
\begin{align}
  D' = \begin{pmatrix}
    \dot{x}_2 & \dot{x}_1 & 0 \\
    1 & 1 & 2 y_{1'} \xi^2 \\
    0 & 0 & \xi
  \end{pmatrix}.
\end{align}
Since $D'$ is not identically singular, the method terminates at this point.

\subsection{Transistor Amplifier}
\label{sec:transistor_amplifier}

Next we consider a transistor amplifier problem arising from an electrical network~\cite{Mazzia2008}.
The problem is described by an index-1 DAE in the following form
\begin{align} \label{eq:transistor_amplifier}
  \left\{\begin{aligned}
    F_1:&& C_1(\dot{x}_1-\dot{x}_2) + (x_1-U_e(t))/R_0 &= 0, \\
    F_2:&& - C_1(\dot{x}_1 - \dot{x}_2) - U_b/R_2 + x_2 (1/R_1 + 1/R_2) - (\alpha - 1)g(x_2 - x_3) &= 0, \\
    F_3:&& C_2 \dot{x}_3 + x_3/R_3 - g(x_2 - x_3) &= 0, \\
    F_4:&& C_3 (\dot{x}_4 - \dot{x}_5) + (x_4 - U_b)/R_4 + \alpha  g(x_2 - x_3) &= 0, \\
    F_5:&& - C_3 (\dot{x}_4 - \dot{x}_5) - U_b/R_6 + x_5 (1/R_5 + 1/R_6) - (\alpha  - 1)g(x_5 - x_6) &= 0, \\
    F_6:&& C_4 \dot{x}_6  + x_6/R_7- g(x_5 - x_6) &= 0, \\
    F_7:&& C_5 (\dot{x}_7 - \dot{x}_8) + (x_7 - U_b)/R_8 + \alpha  g(x_5 - x_6) &= 0, \\
    F_8:&& - C_5 (\dot{x}_7 - \dot{x}_8) + x_8/R_9 &= 0,
  \end{aligned}\right.
\end{align}
where $g(x) = \beta (\exp(x / U_F) - 1)$ and $U_e(t) = 0.1 \sin(200 \pi t)$ with nonzero parameters $U_b$, $U_F$, $\alpha$, $\beta$, $R_0, R_1, \ldots, R_9$, and $C_1, \ldots, C_5$.

A dual optimal solution on~\eqref{eq:transistor_amplifier} is given by $p = (0, \ldots, 0)$ and $q = (1, \ldots, 1) \in \setZ^8$.
The system Jacobian corresponding $(p, q)$ is a singular constant matrix
\begin{align}
  D = \begin{pmatrix}
     C_1 & -C_1 &   0 &    0 &    0 &   0 &    0 &    0 \\
    -C_1 &  C_1 &   0 &    0 &    0 &   0 &    0 &    0 \\
       0 &    0 & C_2 &    0 &    0 &   0 &    0 &    0 \\
       0 &    0 &   0 &  C_3 & -C_3 &   0 &    0 &    0 \\
       0 &    0 &   0 & -C_3 &  C_3 &   0 &    0 &    0 \\
       0 &    0 &   0 &    0 &    0 & C_4 &    0 &    0 \\
       0 &    0 &   0 &    0 &    0 &   0 &  C_5 & -C_5 \\
       0 &    0 &   0 &    0 &    0 &   0 & -C_5 &  C_5
  \end{pmatrix}.
\end{align}
One possible selection of $(r, I, J)$ is $r = 1, I = \set{2}$ and $J = \set{1}$.

On the substitution method, we solve $F_2 = 0$ for $\dot{x}_1$ to get
\begin{align} \label{eq:example_2_solved_for_x1_dot}
  \dot{x}_1 = \dot{x}_2 + (-U_b/R_2 + x_2 (1/R_1 + 1/R_2) - (\alpha - 1)g(x_2 - x_3)) / C_1
\end{align}
and substitute~\eqref{eq:example_2_solved_for_x1_dot} into $F_1 = 0$.
Then the first equation is modified into
\begin{align}
  \sub{F}_1: -U_b/R_2 + x_2 (1/R_1 + 1/R_2) - (\alpha - 1)g(x_2 - x_3) + (x_1-U_e(t))/R_0 = 0
\end{align}
and the dual optimal solution is updated to $p' = (1, 0, 0, 0, 0, 0, 0,1)$ and $q' = q$.
The substitution method modifies the DAE twice more in the same manner for $(r, I, J) = (4, \set{5}, \set{4})$ and $(7, \set{8}, \set{7})$, and outputs the following DAE
\begin{align} \label{eq:transistor_amplifier_sub}
  \left\{\begin{aligned}
    \sub{F}_1:&& - U_b/R_2 + x_2 (1/R_1 + 1/R_2) - (\alpha - 1)g(x_2 - x_3) + (x_1-U_e(t))/R_0 &= 0, \\
    F_2^{\phantom{\mathrm{sub}}}:&& - C_1(\dot{x}_1 - \dot{x}_2) - U_b/R_2 + x_2 (1/R_1 + 1/R_2) - (\alpha - 1)g(x_2 - x_3) &= 0, \\
    F_3^{\phantom{\mathrm{sub}}}:&& C_2 \dot{x}_3 + x_3/R_3 - g(x_2 - x_3) &= 0, \\
    \sub{F}_4:&& - U_b/R_6 + x_5 (1/R_5 + 1/R_6) - (\alpha  - 1)g(x_5 - x_6) \quad \\
    && {} + (x_4 - U_b)/R_4 + \alpha  g(x_2 - x_3) &= 0, \\
    F_5^{\phantom{\mathrm{sub}}}:&& - C_3 (\dot{x}_4 - \dot{x}_5) - U_b/R_6 + x_5 (1/R_5 + 1/R_6) - (\alpha  - 1)g(x_5 - x_6) &= 0, \\
    F_6^{\phantom{\mathrm{sub}}}:&& C_4 \dot{x}_6  + x_6/R_7- g(x_5 - x_6) &= 0, \\
    \sub{F}_7:&& x_8/R_9 + (x_7 - U_b)/R_8 + \alpha  g(x_5 - x_6) &= 0, \\
    F_8^{\phantom{\mathrm{sub}}}:&& - C_5 (\dot{x}_7 - \dot{x}_8) + x_8/R_9 &= 0,
  \end{aligned}\right.
\end{align}
which has a nonsingular system Jacobian.

The augmentation method also modifies the DAE~\eqref{eq:transistor_amplifier} three times for $(r, I, J) = (1, \set{2}, \set{1})$, $(4, \set{5}, \set{4})$ and $(7, \set{8}, \set{7})$.
Due to limitations of space, we just describe the resulting DAE in the following:
\begin{align}
  \left\{\begin{aligned}
    \aug{F}_1:&& C_1(y_1 - \xi_2) + (x_1-U_e(t))/R_0 &= 0, \\
    F_2^{\phantom{\mathrm{aug}}}:&& - C_1(\dot{x}_1 - \dot{x}_2) - U_b/R_2 + x_2 (1/R_1 + 1/R_2) - (\alpha - 1)g(x_2 - x_3) &= 0, \\
    \aug{F}_{2'}:&& - C_1(y_1 - \xi_2) - U_b/R_2 + x_2 (1/R_1 + 1/R_2) - (\alpha - 1)g(x_2 - x_3) &= 0, \\
    F_3^{\phantom{\mathrm{aug}}}:&& C_2 \dot{x}_3 + x_3/R_3 - g(x_2 - x_3) &= 0, \\
    \aug{F}_{4}:&& C_3 (y_4 - \xi_5) + (x_4 - U_b)/R_4 + \alpha  g(x_2 - x_3) &= 0, \\
    F_5^{\phantom{\mathrm{aug}}}:&& - C_3 (\dot{x}_4 - \dot{x}_5) - U_b/R_6 + x_5 (1/R_5 + 1/R_6) - (\alpha  - 1)g(x_5 - x_6) &= 0, \\
    \aug{F}_{5'}:&& - C_3 (y_4 - \xi_5) - U_b/R_6 + x_5 (1/R_5 + 1/R_6) - (\alpha  - 1)g(x_5 - x_6) &= 0, \\
    F_6^{\phantom{\mathrm{aug}}}:&& C_4 \dot{x}_6  + x_6/R_7- g(x_5 - x_6) &= 0, \\
    \aug{F}_7:&& C_5 (y_7 - \xi_8) + (x_7 - U_b)/R_8 + \alpha  g(x_5 - x_6) &= 0, \\
    F_8^{\phantom{\mathrm{aug}}}:&& - C_5 (\dot{x}_7 - \dot{x}_8) + x_8/R_9 &= 0, \\
    \aug{F}_{8'}:&& - C_5 (y_7 - \xi_8) + x_8/R_9 &= 0,
  \end{aligned}\right.
\end{align}
where $y_1, y_4$ and $y_7$ are new variables corresponding to $\dot{x}_1, \dot{x}_4,$ and $\dot{x}_7$, respectively, and $\xi_2, \xi_5$ and $\xi_8$ are arbitrary constants corresponding to $\dot{x}_2, \dot{x}_5$ and $\dot{x}_8$, respectively.

Indeed, the LC-method can also modify the DAE~\eqref{eq:transistor_amplifier} into~\eqref{eq:transistor_amplifier_sub}.
In general, the substitution method and the LC-method return the same DAE under some reasonable restrictions; see the appendix for details.

\section{Numerical Experiments}
\label{sec:numerical_experiments}

We applied our library in practice to the following four DAEs.
The DAEs have identically singular system Jacobian, and thus the MS-method, which is the index reduction method used in MATLAB, cannot be applied to them.

\begin{enumerate}[label={(\alph*)}]
  \item Nonlinearly modified pendulum (index-3):
    \begin{align}
      \left\{\begin{aligned}
        \dot{x}_4 - x_1 x_2 \cos x_3 &= 0, \\
        \dot{x}_5 - {x_2}^2 \cos x_3 \sin x_3 + g &= 0, \\ 
        {x_1}^2 + {x_2}^2 \sin^2 x_3 - 1 &= 0, \\
        \tanh (\dot{x}_1 - x_4) &= 0, \\
        {{}\dot{x}}_2 \sin x_3 + x_2 \dot{x}_3 \cos x_3 - x_5 &= 0
      \end{aligned}\right.
    \end{align}
    with parameter $g = 9.8$.
    This DAE is obtained by nonlinearly changing the variable $(y, z, \lambda, v_y, v_z)$ of a simple pendulum DAE
    \begin{align}
      \left\{\begin{aligned}
        \dot{v}_y - y\lambda &= 0, \\
        \dot{v}_z - z\lambda + g &= 0, \\
        y^2 + z^2 - 1 &= 0, \\
        \dot{y} - v_y &= 0, \\
        \dot{z} - v_z &= 0 \\
      \end{aligned}\right.
    \end{align}
    by $(y, z, \lambda, v_y, v_z) = (x_1, x_2 \sin x_3, x_2 \cos x_3, x_4, x_5)$.
    In addition, we equivalently changed the fourth equation $\dot{x}_1 - x_4 = 0$ to $\tanh (\dot{x}_1 - x_4) = 0$.
  \item Robotic arm (index-5):
    \begin{align}
      \left\{\begin{aligned}
        \ddot{x}_1 - 2 c(x_{3}) (\dot{x}_1 + \dot{x}_3)^2
         - {{}\dot{x}_1}^2 d(x_3) + (x_2 - 2x_3)(a(x_3) + 2 b(x_3)) \quad \\
           {}-{} a(x_3) (x_4 - x_5) &= 0, \\
        \ddot{x}_2 + 2c(x_3)(\dot{x}_1 + \dot{x}_3)^2 + {{}\dot{x}_1}^2 d(x_3) + (x_2 - 2x_3)(1-3a(x_3)-2b(x_3)) \quad \\
          {}+{} a(x_3)(x_4 - x_5) + x_5 &= 0, \\
        \ddot{x}_3 + 2c(x_3)(\dot{x}_1 + \dot{x}_3)^2 + {{}\dot{x}_1}^2 d(x_3) + (x_2 - 2x_3)(a(x_3) - 9b(x_3)) \quad \\
          {}+{} 2\dot{x}_1^2 c(x_3) + d(x_3)(\dot{x}_1 + \dot{x}_3)^2 + (a(x_3) + b(x_3))(x_1 - x_2) &= 0, \\
        \cos x_1 + \cos (x_1+x_3) - p_1(t) &= 0, \\
        \sin x_1 + \sin (x_1+x_3) - p_2(t) &= 0,
      \end{aligned}\right.
    \end{align}
    where
    \begin{align}
      &p_1(t) = \cos(1 - \e^t) + \cos(1 - t),
      \,
      p_2(t) = \sin(1 - \e^t) + \sin(1 - t),
      \\
      &a(s) = \frac{2}{2 - \cos^2 s},
      \,
      b(s) = \frac{\cos s}{2 - \cos^2 s},
      \,
      c(s) = \frac{\sin s}{2 - \cos^2 s},
      \,
      d(s) = \frac{\sin s \cos s}{2 - \cos^2 s}.
    \end{align}
    The robotic arm DAE arises from the path control of a two-link, flexible joint and planar robotic arm~\cite{Campbell1995b}.
    The above formulation is a slightly modified version given in the preliminary paper of~\cite{Tan2017} available on arXiv.
    
  \item Transistor amplifier (index-1): the DAE~\eqref{eq:transistor_amplifier}.
    The values of parameters are shown in the appendix.
  \item Ring modulator (index-2):
    \begin{align}
      \left\{\begin{aligned}
        \dot{x}_1 + (x_1/R - x_{8} + 0.5x_{10} - 0.5x_{11} - x_{14})/C &= 0, \\
        \dot{x}_2 + (x_{2}/R - x_{9} + 0.5x_{12} - 0.5x_{13} - x_{15})/C &= 0, \\
        x_{10} - q(U_{D1}) + q(U_{D4}) &= 0, \\
        x_{11} - q(U_{D2}) + q(U_{D3}) &= 0, \\
        x_{12} + q(U_{D1}) - q(U_{D3}) &= 0, \\
        x_{13} + q(U_{D2}) - q(U_{D4})  &= 0, \\
        \dot{x}_7 + (x_{7}/R_p - q(U_{D1}) - q(U_{D2})  + q(U_{D3}) + q(U_{D4}))/C_p &= 0, \\
        \dot{x}_8 + x_{1}/L_h &= 0, \\
        \dot{x}_9 + x_{2}/L_h &= 0, \\
        \dot{x}_{10} + (-0.5x_1 + x_3 + R_{g2} x_{10})/L_{s2} &= 0, \\
        \dot{x}_{11} + (0.5x_1 - x_4 + R_{g3} x_{11})/L_{s3} &= 0, \\
        \dot{x}_{12} + (-0.5x_2 + x_5 + R_{g2} x_{12})/L_{s2} &= 0, \\
        \dot{x}_{13} + (0.5x_2 - x_6 + R_{g3} x_{13})/L_{s3} &= 0, \\
        \dot{x}_{14} + (x_1 + (R_{g1} + R_i)x_{14} - U_{\mathrm{in1}}(t) )/L_{s1} &= 0, \\
        \dot{x}_{15} + (x_2 + (R_c + R_{g1})x_{15} )/L_{s1} &= 0,
      \end{aligned}\right.
    \end{align}
    where
    \begin{align}
      &U_{D1} = x_3 - x_5 - x_7 - U_{\mathrm{in2}}(t),
      \quad
      U_{D2} = -x_4 + x_6 - x_7 - U_{\mathrm{in2}}(t),
      \\&
      U_{D3} =  x_4 + x_5 + x_7 + U_{\mathrm{in2}}(t),
      \quad
      U_{D4} = -x_3 - x_6 + x_7 + U_{\mathrm{in2}}(t),
      \\&
      q(U) = \gamma \prn{\e^{\delta U} - 1},
      \quad
      U_{\mathrm{in1}}(t) = 0.5 \sin 2000 \pi t,
      \quad
      U_{\mathrm{in2}}(t) = 2 \sin 20000 \pi t
    \end{align}
    with parameters $C, C_p, L_h, L_{s1}, L_{s2}, L_{s3}, \gamma, \delta, R, R_p, R_{g1}, R_{g2}, R_{g3}, R_i$ and $R_c$.
    Their numerical values are given in the appendix.
    The DAE represents an electrical network describing the behaviour of a ring modulator~\cite{Mazzia2008}.
    The above formulation is obtained by setting $C_s = 0$ in the original problem.
\end{enumerate}

We computed numerical solutions of DAEs (a)--(d) after performing the following three kinds of preprocessing: (i) no index reduction, (ii) reduce the index by the MS-method after applying the substitution method, and (iii) reduce the index after using the augmentation method; what the default DAE solver in MATLAB can do for the DAEs is only (i).
For the rank computation of system Jacobian and the process of finding $(r, I, J)$, we adopted the fast symbolic Gaussian elimination algorithm by Sasaki--Murao~\cite{Sasaki1982}.
The numerical solutions were computed by \texttt{ode15i} in MATLAB, which is a variable-step variable-order (VSVO) solver for index-0 or 1 DAEs based on the backward differentiation formulas (BDFs)~\cite{Shampine2002}.
The parameters of \texttt{ode15i} were set to the default values: $\texttt{AbsTol} = 10^{-6}$ and $\texttt{RelTol} = 10^{-3}$.
We explicitly provided Jacobian matrices to \texttt{ode15i} through the \texttt{Jacobian} option.
All the computation were performed on MATLAB R2019a.

\paragraph{Numerical Results.}
\begin{figure}[htbp]
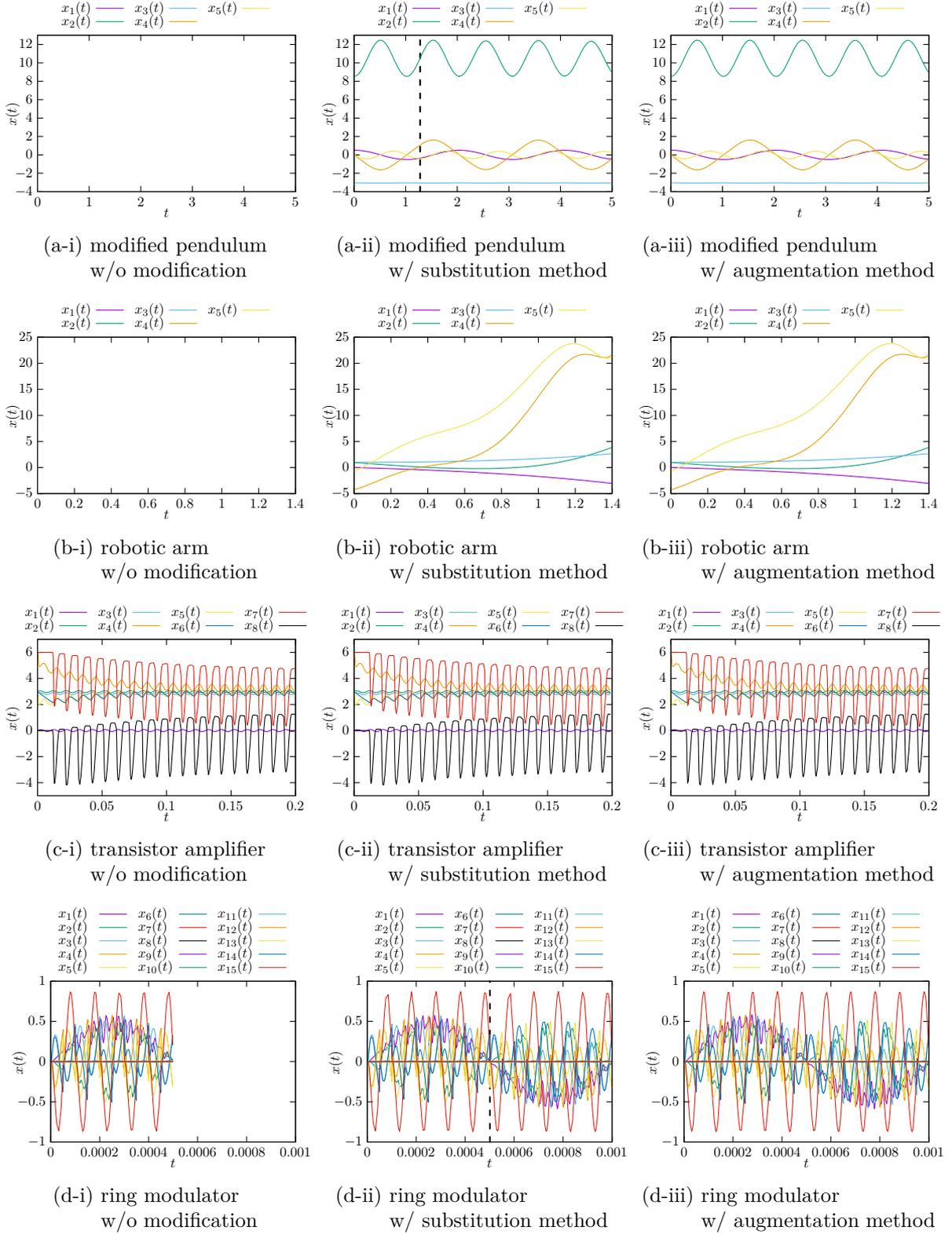

  \iftrue
  \begin{minipage}[b]{0.333\hsize}%
    \centering
    \begin{tikzpicture}[gnuplot]
\tikzset{every node/.append style={scale=0.60}}
\path (0.000,0.000) rectangle (5.330,4.000);
\gpcolor{color=gp lt color border}
\gpsetlinetype{gp lt border}
\gpsetdashtype{gp dt solid}
\gpsetlinewidth{1.00}
\draw[gp path] (0.680,0.499)--(0.770,0.499);
\draw[gp path] (4.999,0.499)--(4.909,0.499);
\node[gp node right] at (0.680,0.499) {$-4$};
\draw[gp path] (0.680,0.810)--(0.770,0.810);
\draw[gp path] (4.999,0.810)--(4.909,0.810);
\node[gp node right] at (0.680,0.810) {$-2$};
\draw[gp path] (0.680,1.120)--(0.770,1.120);
\draw[gp path] (4.999,1.120)--(4.909,1.120);
\node[gp node right] at (0.680,1.120) {$0$};
\draw[gp path] (0.680,1.431)--(0.770,1.431);
\draw[gp path] (4.999,1.431)--(4.909,1.431);
\node[gp node right] at (0.680,1.431) {$2$};
\draw[gp path] (0.680,1.741)--(0.770,1.741);
\draw[gp path] (4.999,1.741)--(4.909,1.741);
\node[gp node right] at (0.680,1.741) {$4$};
\draw[gp path] (0.680,2.052)--(0.770,2.052);
\draw[gp path] (4.999,2.052)--(4.909,2.052);
\node[gp node right] at (0.680,2.052) {$6$};
\draw[gp path] (0.680,2.363)--(0.770,2.363);
\draw[gp path] (4.999,2.363)--(4.909,2.363);
\node[gp node right] at (0.680,2.363) {$8$};
\draw[gp path] (0.680,2.673)--(0.770,2.673);
\draw[gp path] (4.999,2.673)--(4.909,2.673);
\node[gp node right] at (0.680,2.673) {$10$};
\draw[gp path] (0.680,2.984)--(0.770,2.984);
\draw[gp path] (4.999,2.984)--(4.909,2.984);
\node[gp node right] at (0.680,2.984) {$12$};
\draw[gp path] (0.680,0.499)--(0.680,0.589);
\draw[gp path] (0.680,3.139)--(0.680,3.049);
\node[gp node center] at (0.680,0.314) {$0$};
\draw[gp path] (1.544,0.499)--(1.544,0.589);
\draw[gp path] (1.544,3.139)--(1.544,3.049);
\node[gp node center] at (1.544,0.314) {$1$};
\draw[gp path] (2.408,0.499)--(2.408,0.589);
\draw[gp path] (2.408,3.139)--(2.408,3.049);
\node[gp node center] at (2.408,0.314) {$2$};
\draw[gp path] (3.271,0.499)--(3.271,0.589);
\draw[gp path] (3.271,3.139)--(3.271,3.049);
\node[gp node center] at (3.271,0.314) {$3$};
\draw[gp path] (4.135,0.499)--(4.135,0.589);
\draw[gp path] (4.135,3.139)--(4.135,3.049);
\node[gp node center] at (4.135,0.314) {$4$};
\draw[gp path] (4.999,0.499)--(4.999,0.589);
\draw[gp path] (4.999,3.139)--(4.999,3.049);
\node[gp node center] at (4.999,0.314) {$5$};
\draw[gp path] (0.680,3.139)--(0.680,0.499)--(4.999,0.499)--(4.999,3.139)--cycle;
\node[gp node center,rotate=-270] at (0.275,1.819) {$x(t)$};
\node[gp node center] at (2.839,0.129) {$t$};
\node[gp node left] at (1.002,3.595) {$x_1(t)$};
\gpcolor{rgb color={0.580,0.000,0.827}}
\draw[gp path] (1.662,3.595)--(2.117,3.595);
\gpcolor{color=gp lt color border}
\node[gp node left] at (1.002,3.370) {$x_2(t)$};
\gpcolor{rgb color={0.000,0.620,0.451}}
\draw[gp path] (1.662,3.370)--(2.117,3.370);
\gpcolor{color=gp lt color border}
\node[gp node left] at (2.227,3.595) {$x_3(t)$};
\gpcolor{rgb color={0.337,0.706,0.914}}
\draw[gp path] (2.887,3.595)--(3.342,3.595);
\gpcolor{color=gp lt color border}
\node[gp node left] at (2.227,3.370) {$x_4(t)$};
\gpcolor{rgb color={0.902,0.624,0.000}}
\draw[gp path] (2.887,3.370)--(3.342,3.370);
\gpcolor{color=gp lt color border}
\node[gp node left] at (3.452,3.595) {$x_5(t)$};
\gpcolor{rgb color={0.941,0.894,0.259}}
\draw[gp path] (4.112,3.595)--(4.567,3.595);
\gpcolor{color=gp lt color border}
\draw[gp path] (0.680,3.139)--(0.680,0.499)--(4.999,0.499)--(4.999,3.139)--cycle;
\gpdefrectangularnode{gp plot 1}{\pgfpoint{0.680cm}{0.499cm}}{\pgfpoint{4.999cm}{3.139cm}}
\end{tikzpicture}
%
    \subcaption{\tabular[t]{@{}l@{}}modified pendulum\\w/o modification\endtabular}
  \end{minipage}%
  \begin{minipage}[b]{0.333\hsize}%
    \centering
    \input{pendulum_sub.tex}%
    \subcaption{\tabular[t]{@{}l@{}}modified pendulum\\w/ substitution method\endtabular}
  \end{minipage}%
  \begin{minipage}[b]{0.333\hsize}%
    \centering
    \input{pendulum_aug.tex}%
    \subcaption{\tabular[t]{@{}l@{}}modified pendulum\\w/ augmentation method\endtabular}
  \end{minipage}
  \begin{minipage}[b]{0.333\hsize}%
    \centering
    \begin{tikzpicture}[gnuplot]
\tikzset{every node/.append style={scale=0.60}}
\path (0.000,0.000) rectangle (5.330,4.000);
\gpcolor{color=gp lt color border}
\gpsetlinetype{gp lt border}
\gpsetdashtype{gp dt solid}
\gpsetlinewidth{1.00}
\draw[gp path] (0.680,0.499)--(0.770,0.499);
\draw[gp path] (4.999,0.499)--(4.909,0.499);
\node[gp node right] at (0.680,0.499) {$-5$};
\draw[gp path] (0.680,0.939)--(0.770,0.939);
\draw[gp path] (4.999,0.939)--(4.909,0.939);
\node[gp node right] at (0.680,0.939) {$0$};
\draw[gp path] (0.680,1.379)--(0.770,1.379);
\draw[gp path] (4.999,1.379)--(4.909,1.379);
\node[gp node right] at (0.680,1.379) {$5$};
\draw[gp path] (0.680,1.819)--(0.770,1.819);
\draw[gp path] (4.999,1.819)--(4.909,1.819);
\node[gp node right] at (0.680,1.819) {$10$};
\draw[gp path] (0.680,2.259)--(0.770,2.259);
\draw[gp path] (4.999,2.259)--(4.909,2.259);
\node[gp node right] at (0.680,2.259) {$15$};
\draw[gp path] (0.680,2.699)--(0.770,2.699);
\draw[gp path] (4.999,2.699)--(4.909,2.699);
\node[gp node right] at (0.680,2.699) {$20$};
\draw[gp path] (0.680,3.139)--(0.770,3.139);
\draw[gp path] (4.999,3.139)--(4.909,3.139);
\node[gp node right] at (0.680,3.139) {$25$};
\draw[gp path] (0.680,0.499)--(0.680,0.589);
\draw[gp path] (0.680,3.139)--(0.680,3.049);
\node[gp node center] at (0.680,0.314) {$0$};
\draw[gp path] (1.297,0.499)--(1.297,0.589);
\draw[gp path] (1.297,3.139)--(1.297,3.049);
\node[gp node center] at (1.297,0.314) {$0.2$};
\draw[gp path] (1.914,0.499)--(1.914,0.589);
\draw[gp path] (1.914,3.139)--(1.914,3.049);
\node[gp node center] at (1.914,0.314) {$0.4$};
\draw[gp path] (2.531,0.499)--(2.531,0.589);
\draw[gp path] (2.531,3.139)--(2.531,3.049);
\node[gp node center] at (2.531,0.314) {$0.6$};
\draw[gp path] (3.148,0.499)--(3.148,0.589);
\draw[gp path] (3.148,3.139)--(3.148,3.049);
\node[gp node center] at (3.148,0.314) {$0.8$};
\draw[gp path] (3.765,0.499)--(3.765,0.589);
\draw[gp path] (3.765,3.139)--(3.765,3.049);
\node[gp node center] at (3.765,0.314) {$1$};
\draw[gp path] (4.382,0.499)--(4.382,0.589);
\draw[gp path] (4.382,3.139)--(4.382,3.049);
\node[gp node center] at (4.382,0.314) {$1.2$};
\draw[gp path] (4.999,0.499)--(4.999,0.589);
\draw[gp path] (4.999,3.139)--(4.999,3.049);
\node[gp node center] at (4.999,0.314) {$1.4$};
\draw[gp path] (0.680,3.139)--(0.680,0.499)--(4.999,0.499)--(4.999,3.139)--cycle;
\node[gp node center,rotate=-270] at (0.275,1.819) {$x(t)$};
\node[gp node center] at (2.839,0.129) {$t$};
\node[gp node left] at (1.002,3.595) {$x_1(t)$};
\gpcolor{rgb color={0.580,0.000,0.827}}
\draw[gp path] (1.662,3.595)--(2.117,3.595);
\gpcolor{color=gp lt color border}
\node[gp node left] at (1.002,3.370) {$x_2(t)$};
\gpcolor{rgb color={0.000,0.620,0.451}}
\draw[gp path] (1.662,3.370)--(2.117,3.370);
\gpcolor{color=gp lt color border}
\node[gp node left] at (2.227,3.595) {$x_3(t)$};
\gpcolor{rgb color={0.337,0.706,0.914}}
\draw[gp path] (2.887,3.595)--(3.342,3.595);
\gpcolor{color=gp lt color border}
\node[gp node left] at (2.227,3.370) {$x_4(t)$};
\gpcolor{rgb color={0.902,0.624,0.000}}
\draw[gp path] (2.887,3.370)--(3.342,3.370);
\gpcolor{color=gp lt color border}
\node[gp node left] at (3.452,3.595) {$x_5(t)$};
\gpcolor{rgb color={0.941,0.894,0.259}}
\draw[gp path] (4.112,3.595)--(4.567,3.595);
\gpcolor{color=gp lt color border}
\draw[gp path] (0.680,3.139)--(0.680,0.499)--(4.999,0.499)--(4.999,3.139)--cycle;
\gpdefrectangularnode{gp plot 1}{\pgfpoint{0.680cm}{0.499cm}}{\pgfpoint{4.999cm}{3.139cm}}
\end{tikzpicture}
%
    \subcaption{\tabular[t]{@{}l@{}}robotic arm\\w/o modification\endtabular}
  \end{minipage}%
  \begin{minipage}[b]{0.333\hsize}%
    \centering
    \input{robot_sub.tex}%
    \subcaption{\tabular[t]{@{}l@{}}robotic arm\\w/ substitution method\endtabular}
  \end{minipage}%
  \begin{minipage}[b]{0.333\hsize}%
    \centering
    \input{robot_aug.tex}%
    \subcaption{\tabular[t]{@{}l@{}}robotic arm\\w/ augmentation method\endtabular}
  \end{minipage}
  \begin{minipage}[b]{0.333\hsize}%
    \centering
    \input{transamp_none.tex}%
    \subcaption{\tabular[t]{@{}l@{}}transistor amplifier\\w/o modification\endtabular}
  \end{minipage}%
  \begin{minipage}[b]{0.333\hsize}%
    \centering
    \input{transamp_sub.tex}%
    \subcaption{\tabular[t]{@{}l@{}}transistor amplifier\\w/ substitution method\endtabular}
  \end{minipage}%
  \begin{minipage}[b]{0.333\hsize}%
    \centering
    \input{transamp_aug.tex}%
    \subcaption{\tabular[t]{@{}l@{}}transistor amplifier\\w/ augmentation method\endtabular}
  \end{minipage}
  \begin{minipage}[b]{0.333\hsize}%
    \centering
    \input{ringmod_none.tex}%
    \subcaption{\tabular[t]{@{}l@{}}ring modulator\\w/o modification\endtabular}
  \end{minipage}%
  \begin{minipage}[b]{0.333\hsize}%
    \centering
    \input{ringmod_sub.tex}%
    \subcaption{\tabular[t]{@{}l@{}}ring modulator\\w/ substitution method\endtabular}
  \end{minipage}%
  \begin{minipage}[b]{0.333\hsize}%
    \centering
    \input{ringmod_aug.tex}%
    \subcaption{\tabular[t]{@{}l@{}}ring modulator\\w/ augmentation method\endtabular}
  \end{minipage}
  \fi
  \caption{%
    Numerical solutions of the experiments.
    In the $4 \times 3$ array of graphs, each row corresponds to a DAE and each column corresponds to a modification method.
    The dotted lines in (a-ii) and (d-ii) indicate that the DAEs were re-modified at those times.
  }
  \label{fig:numerical_solutions}
\end{figure}
\cref{fig:numerical_solutions} shows numerical solutions obtained in the experiments.
Without index reduction, \texttt{ode15i} could not yield a numerical solution of the index-3 DAE (a) or the index-5 DAE (b).
As for the index-2 DAE (d), \texttt{ode15i} first output a numerical solution but stopped at $t \approx 0.00045$.
These results indicate that it is difficult for \texttt{ode15i} to solve high-index DAEs stably.

Both the substitution method and the augmentation method successfully modified all the DAEs (a)--(d).
With the preprocessing by the substitution method, \texttt{ode15i} had stopped at $t \approx 1.278810$ and $t \approx 0.000501$ for (a) and (d), respectively.
This is because our methods ensure the equivalence of DAEs only in an area in which $D[I, J]$ is identically nonsingular.
Nevertheless, by re-modifying the DAE at that time, our program could go on the computation.
This process, called the \emph{dynamic pivoting}, can be applied by re-choosing $(r, I, J)$ such that $D[I, J]$ is far from singular at the present point.
The dynamic pivoting is also known to be needed in the MS-method~\cite{Mattsson1993}.

In the first application of the substitution method to the DAE (b), we used the following values of $(p, q, r, I, J)$:
\begin{align}
  p &= (0,0,1,1,1,1,0,0,0,0,0,0,0,0,0),\\
  q &= (1,1,1,1,1,1,1,1,1,1,1,1,1,1,1), \label{eq:good_pqrIJ} \\
  r &= 11,\, I = \set{3,4,5,6,10,12,13},\, J = \set{3,5,6,10,11,12,13}.
\end{align}
There exists another possible values of $(p, q, r, I, J)$ as follows:
\begin{align}
  p &= (0,0,0,0,0,0,0,0,0,0,0,0,0,0,0),\\
  q &= (1,1,0,0,0,0,1,1,1,1,1,1,1,1,1), \label{eq:bad_pqrIJ} \\
  r &= 5,\, I = \set{3,4,6},\, J = \set{3,4,5}.
\end{align}
The values in~\eqref{eq:bad_pqrIJ} seem to be superior to~\eqref{eq:good_pqrIJ} in that the substitution method need to differentiate equations for~\eqref{eq:good_pqrIJ} but not for~\eqref{eq:bad_pqrIJ}, and the size $\card{I}$ of the equation system~\eqref{eq:F_I_diff} for~\eqref{eq:bad_pqrIJ} is smaller than that of~\eqref{eq:good_pqrIJ}.
However, the equation-solving routine in MATLAB could not return a solution of the equation system~\eqref{eq:F_I_diff} for~\eqref{eq:bad_pqrIJ}.
This is because the substitution method with~\eqref{eq:bad_pqrIJ} need to solve the following equation system
\begin{align}
  \left\{\begin{aligned}
    x_{10} - \gamma\e^{\delta (x_3 - x_5 - x_7 - U_{\mathrm{in2}}(t))} + \gamma\e^{\delta (-x_3 - x_6 + x_7 + U_{\mathrm{in2}}(t))} &= 0, \\
    x_{11} - \gamma\e^{\delta (-x_4 + x_6 - x_7 - U_{\mathrm{in2}}(t))} + \gamma\e^{\delta (x_4 + x_5 + x_7 + U_{\mathrm{in2}}(t))} &= 0, \\
    x_{13} + \gamma\e^{\delta (-x_4 + x_6 - x_7 - U_{\mathrm{in2}}(t))} - \gamma\e^{\delta (-x_3 - x_6 + x_7 + U_{\mathrm{in2}}(t))}  &= 0
  \end{aligned}\right.
\end{align}
for $x_3$, $x_4$ and $x_6$, while the solution cannot be represented by a combination of the elementary functions.
On the other hand, the equation system~\eqref{eq:F_I_diff} for~\eqref{eq:good_pqrIJ} is linear because $\dot{x}_3, \dot{x}_5$ and $\dot{x}_6$ appear linearly in the differentiation of the 3--6th equations in (d).
Thus the substitution method works with~\eqref{eq:good_pqrIJ} but not with~\eqref{eq:bad_pqrIJ}.
As we have seen, the substitution method may not run depending on the choice of $(p, q, r, I, J)$.
The experimental results show that the augmentation method successfully serves as a remedy for this issue.

\section{Conclusion}
\label{sec:conclusion}

In this paper, we have presented two modification methods for nonlinear DAEs, called the substitution method and the augmentation method.
Based on the combinatorial relaxation approach, both methods modify DAEs into other DAEs for which the structural preprocessing methods work.
The substitution method modifies DAEs using the IFT and has a merit that it retains the size of DAEs.
The augmentation method modifies DAEs by appending new variables and equations, and is advantageous in that it does not require an equation-solving engine and keeps DAEs' sparsity.
The numerical experiments have shown that both methods can modify DAEs that the standard DAE solver in MATLAB cannot handle into amenable forms.
While the success of the substitution method depends on the selection of the values of $(p, q, r, I, J)$, the augmentation method has successfully served as a remedy for this problem.
Numerical experiments on sparse DAEs are left for further investigation.

\section*{Acknowledgments}
The author thanks Satoru Iwata for careful reading and helpful comments, and thanks Mizuyo Takamatsu for discussions.
This work was supported by JST CREST Grant Number JPMJCR14D2, Japan, Grant-in-Aid for JSPS Research Fellow Grant Number JP18J22141, Japan and JST ACT-I Grant Number JPMJPR18U9, Japan.


\bibliographystyle{abbrv}
\bibliography{references}

\appendix
\section{Appendix}

\subsection{Relation between Substitution Method and LC-method}
\label{sec:relation_to_lc_method}

We discuss a relation between the substitution method and the LC-method described in \cref{sec:the_LC-method}.
One may conjecture that the sets of possible outputs of these two methods are the same if a DAE can be modified by both methods.
However, this is false because the cokernel of the system Jacobian contains infinitely many vectors and the LC-method returns different DAEs for different cokernel vectors, while the number of possible $(r, I, J)$ in the substitution method is finite.
Indeed, the conjecture is true if we restrict the selection of the cokernel vector as follows.

\begin{theorem} \label{thm:relation_to_LC}
  Consider a DAE~\eqref{eq:dae} satisfying \textup{(I1)--(I3)}.
  Let $(r, I, J)$ be a triple satisfying \textup{(C1)--(C3)} and $D$ denote the system Jacobian of $F$ with respect to a dual optimal solution $(p, q)$.
  Define a row vector $u = (u_i(t, x, \dot{x}, \ldots, x^{(l)}))_{i \in R}$ by
  \begin{align}
    u_i \defeq \begin{cases}
        1 & (i = r), \\
        -D[\set{r}, J] \prn{\prn{D[I, J]}^{-1}[J, \set{i}]} & (i \in I), \\
        0 & (\text{otherwise})
    \end{cases}
  \end{align}
  for $i \in R$, where $\prn{D[I, J]}^{-1}[J, \set{i}]$ is the $i$-th column vector in the inverse of $D[I, J]$.
  Then $u$ is a cokernel vector of $D$.
  In addition, if $u$ satisfies the validity condition~\eqref{cond:LC} of the LC-method, then the output $\sub{F}$ of the substitution method is the same as the output $\LC{F}$ of the LC-method with respect to $u$.
  If $u$ does not satisfy~\eqref{cond:LC}, neither does any cokernel vector $v$ of $D$ with $\supp v = I \cup \set{r}$.
\end{theorem}

\begin{proof}
  We first show that $u$ is a cokernel vector of $D$.
  From the definition of $u$, it holds
  \begin{align} \label{eq:thm8.1_uDRJ}
    uD[R, J]
    = D[\set{r}, J] - D[\set{r}, J] \prn{D[I, J]}^{-1} D[I, J] = D[\set{r}, J] - D[\set{r}, J] = 0
  \end{align}
  and
  \begin{align} \label{eq:thm8.1}
    uD[R, T]
    = D[\set{r}, T] - D[\set{r}, J] \prn{D[I, J]}^{-1} D[I, T],
  \end{align}
  where $T \defeq C \setminus J$.
  Here, the right-hand side of~\eqref{eq:thm8.1} is the Schur complement of $D[I, J]$ in the following block matrix
  \begin{align}
    D[I \cup \set{r}, C]
    = \begin{pmatrix}
      D[\set{r}, J] & D[\set{r}, T] \\
      D[I, J]       & D[I, T] \\
    \end{pmatrix}.
  \end{align}
  Since $\rank D[I \cup \set{r}, C] = \rank D[I, J]$ from the conditions (C1) and (C2), the rank of $uD[R, T]$ must be zero, which means $uD[R, T] = 0$.
  Thus $u$ is a cokernel vector of $D$ by this and~\eqref{eq:thm8.1_uDRJ}.

  Next suppose that $u$ satisfies the validity condition~\eqref{cond:LC} of the LC-method.
  Let $u_I \defeq (u_i)_{i \in I}$.
  Then the modified function $\LC{F}_r$ in~\eqref{eq:LC_F_r} can be expressed as $\LC{F}_r = F_r + u_I F_I^{(p_I - p_r \onevec)}$, where we used $u_r = 1$ in the present setting.
  Let $\mathcal{C}$ and $\mathcal{J}$ be index sets defined in~\eqref{def:mathcal_C} and~\eqref{def:mathcal_J}, respectively.
  In order to show $\sub{F}_r = \LC{F}_r$, we prove
  \begin{align} \label{eq:thm8.1_2}
    \pdif{\sub{F}_r}{x_j^{(k)}}(t, X_{\mathcal{C} \setminus \mathcal{J}}) = \pdif{\LC{F}_r}{x_j^{(k)}}(t, X)
  \end{align}
  for all $j \in C,$ a nonnegative integer $k$ and a point $(t, X)$ in the domain $\sub{\mathbb{T}} \times \sub{\Omega}$ of $\sub{F}$.
  If~\eqref{eq:thm8.1_2} holds, then the difference of $\sub{F}_r$ and $\LC{F}_r$ is a function depending only on $t$.
  Now since $\sub{F} = 0$ and $\LC{F} = 0$ share the same solution set, the difference must be identically zero and we are done.

  We show~\eqref{eq:thm8.1_2}.
  Let denote $(t, X_{\mathcal{C} \setminus \mathcal{J}}, \phi(t, X_{\mathcal{C} \setminus \mathcal{J}}))$ by $A_{t, X}$ for short, where $\phi$ is the explicit function given by~\eqref{eq:phi}.
  Then by the same calculation as~\eqref{eq:schur}, it holds
  \begin{align}
    \pdif{\sub{F}_r}{x_j^{(k)}}(t, X_{\mathcal{C} \setminus \mathcal{J}})
    =
    \pdif{F_r}{x_j^{(k)}}(A_{t, X}) - \pdif{F_r^{(p_r)}}{x_J^{(q)}}(A_{t, X}) \prn{\pdif{F_I^{(p)}}{x_J^{(q)}}(A_{t, X})}^{-1} \pdif{F_I^{(p - p_r \onevec)}}{x_j^{(k)}}(A_{t, X}).
  \end{align}
  On the other hand, the right-hand side of~\eqref{eq:thm8.1_2} is calculated as
  \begin{align}
    \pdif{\LC{F}_r}{x_j^{(k)}}(t, X)
    &=
    \pdif{}{x_j^{(k)}} \restr{\prn{F_r + u_I F_I^{(p - p_r \onevec)}}}{(t, X)} \\
    &=
    \pdif{F_r}{x_j^{(k)}}(t, X) + \pdif{u_I}{x_j^{(k)}}(t, X) F_I^{(p - p_r \onevec)}(t, X) + u_I(t, X) \pdif{F_I^{(p - p_r \onevec)}}{x_j^{(k)}}(t, X)
  \end{align}
  for all $(t, X)$.
  Now since $\LC{F}_r$ does not depend on $X_{\mathcal{J}} = \prn{x_j^{(q_j)}}_{j \in J}$ by~\cite[Lemma~4.2]{Tan2017}, the both sides of the above equation also does not depend on $X_{\mathcal{J}}$.
  By replacing $X_{\mathcal{J}}$ in $X$ with $\phi(t, X_{\mathcal{C} \setminus \mathcal{J}})$, we have
  \begin{align}
    \pdif{\LC{F}_r}{x_j^{(k)}}(t, X)
    &= \pdif{F_r}{x_j^{(k)}}(A_{t, X}) + \pdif{u_I}{x_j^{(k)}}(A_{t, X}) F_I^{(p - p_r \onevec)}(A_{t, X}) + u_I(A_{t, X}) \pdif{F_I^{(p - p_r \onevec)}}{x_j^{(k)}}(A_{t, X}) \\
    &= \pdif{F_r}{x_j^{(k)}}(A_{t, X}) + u_I(A_{t, X}) \pdif{F_I^{(p - p_r \onevec)}}{x_j^{(k)}}(A_{t, X}) \\
    &=
    \pdif{F_r}{x_j^{(k)}}(A_{t, X}) - \pdif{F_r^{(p_r)}}{x_J^{(q)}}(A_{t, X}) \prn{\pdif{F_I^{(p)}}{x_J^{(q)}}(A_{t, X})}^{-1} \pdif{F_I^{(p - p_r \onevec)}}{x_j^{(k)}}(A_{t, X}),
  \end{align}
  where we used $F_I^{(p - p_r \onevec)}(A_{t, X}) = 0$ in the second equality.
  Thus~\eqref{eq:thm8.1_2} holds.

  Finally, suppose that $u$ does not satisfy~\eqref{cond:LC}.
  Let $v = (v_i)_{i \in R}$ be a nonzero cokernel vector of $D$ with $\supp v = I \cup \set{r}$.
  This implies that $(v_i)_{i \in I \cup \set{r}}$ is in the cokernel of $D[I \cup \set{r}, C]$.
  From $\rank D[I \cup \set{r}, C] = \card{I}$, the dimension of the cokernel of $D[I \cup \set{r}, C]$ is 1.
  Thus all cokernel vectors of $D$ having support $I \cup \set{r}$ are multiples of $u$.
  Indeed, it holds $v = v_r u$ from $u_r = 1$.
  Since $u$ does not satisfy~\eqref{cond:LC}, there exists $i \in R$ and $j \in C$ such that $\ord{u_i}{x_j} \geq q_j - p_r$.
  If $\ord{v_r}{x_j} < q_j - p_r$, then $\ord{v_i}{x_j} = \ord{v_r u_i}{x_j} = \ord{u_i}{x_j} \geq q_j - p_r$, which implies that $v$ does not satisfy the validity condition~\eqref{cond:LC}.
  Otherwise, $\ord{v_r}{x_j} \geq q_j - p_r$, which also means that $v$ does not meet~\eqref{cond:LC}.
  \qed
\end{proof}

\subsection{Parameter and Initial Value Settings}

The parameter and initial value settings used in the numerical experiments in \cref{sec:numerical_experiments} are described.
\cref{tbl:param_transamp,tbl:param_ringmod} show the parameter settings of the transistor amplifier DAE and the ring modulator DAE, respectively.
\cref{tbl:initval_pendulum,tbl:initval_robot,tbl:initval_transamp,tbl:initval_ringmod} show the initial value settings of the nonlinearly modified pendulum DAE, the robotic arm DAE, the transistor amplifier DAE and the ring modulator DAE, respectively.

\begin{table}[ht]
  \centering
  \caption{Initial values of the nonlinearly modified pendulum DAE}
  \label{tbl:initval_pendulum}
  \begin{tabular}{crr}
    \hline
    $j$ & \multicolumn{1}{c}{$x_j(0)$} & \multicolumn{1}{c}{$\dot{x}_j(0)$} \\ \hline \hline 
    1 &             0.5 &                  0 \\
    2 & 8.5311195044981 &                  0 \\
    3 & 3.2432815053528 &                  0 \\
    4 &               0 & $-4.2435244785437$ \\
    5 &               0 & $-2.45$            \\\hline
  \end{tabular}
\end{table}

\begin{table}[ht]
  \centering
  \caption{Initial values of the robotic arm DAE}
  \label{tbl:initval_robot}
  \begin{tabular}{crrr}
    \hline
    $j$ & \multicolumn{1}{c}{$x_j(0)$} & \multicolumn{1}{c}{$\dot{x}_j(0)$} & \multicolumn{1}{c}{$\ddot{x}_j(0)$} \\ \hline \hline 
    1 &                  0 &               $-1$ &                   $-1$ \\
    2 &    0.9537503511807 & $-2.5319168790105$ & $-1.147631091390737$ \\
    3 &                  1 &                  0 &                    1 \\
    4 & $-4.2781254864526$ &   10.7800085515996 &     56.1923974325182 \\
    5 & $-0.7437526892114$ &   15.9886113811556 &     62.7105238752326 \\\hline
  \end{tabular}
\end{table}

\begin{table}[ht]
  \centering
  \caption{Parameters of the transistor amplifier DAE}
  \label{tbl:param_transamp}
  \begin{tabular}{cr|cr}
    \hline
    parameter & \multicolumn{1}{c|}{value}   & parameter & \multicolumn{1}{c}{value} \\ \hline \hline 
    $U_b$    & 6         & $R_0$                          & $1000$             \\
    $U_F$    & 0.026     & $R_k \quad (k = 1, \ldots, 9)$ & $9000$             \\
    $\alpha$ & 0.99      & $C_k \quad (k = 1, \ldots, 5)$ & $k \times 10^{-6}$ \\
    $\beta$  & $10^{-6}$ &                                &                   \\\hline
  \end{tabular}
\end{table}

\begin{table}[ht]
  \centering
  \caption{Initial values of the transistor amplifier DAE}
  \label{tbl:initval_transamp}
  \begin{tabular}{crr|crr}
    \hline
    $j$ & \multicolumn{1}{c}{$x_j(0)$} & \multicolumn{1}{c|}{$\dot{x}_j(0)$} & $j$ & \multicolumn{1}{c}{$x_j(0)$} & \multicolumn{1}{c}{$\dot{x}_j(0)$} \\ \hline \hline 
    1 & 0 &    51.3392765171807 & 5 & 3 & $-24.9703285154063$ \\
    2 & 3 &    51.3392765171807 & 6 & 3 & $-83.3333333333333$ \\
    3 & 3 & $-166.666666666667$ & 7 & 6 & $-10.0002764024563$ \\
    4 & 6 & $-24.9703285154063$ & 8 & 0 & $-10.0002764024563$ \\\hline
  \end{tabular}
\end{table}

\begin{table}[ht]
  \centering
  \caption{Parameters of the ring modulator DAE}
  \label{tbl:param_ringmod}
  \begin{tabular}{cr|cr}
    \hline
    parameter & \multicolumn{1}{c|}{value}   & parameter & \multicolumn{1}{c}{value} \\ \hline \hline 
    $C$       & $1.6 \times 10^{-8}$         & $R$       & $25000$ \\
    $C_p$     & $10^{-8}$                    & $R_p$     & $50$    \\
    $L_h$     & 4.45                         & $R_{g1}$  & $36.3$  \\
    $L_{s1}$  & $2 \times 10^{-3}$           & $R_{g2}$  & $17.3$  \\
    $L_{s2}$  & $5 \times 10^{-4}$           & $R_{g3}$  & $17.3$  \\
    $L_{s3}$  & $5 \times 10^{-4}$           & $R_i$     & $50$    \\
    $\gamma$  & $40.67286402 \times 10^{-9}$ & $R_c$     & $600$   \\
    $\delta$  & $17.7493332$                 &           &         \\\hline
  \end{tabular}
\end{table}

\begin{table}[ht]
  \centering
  \caption{Initial values of the ring modulator DAE}
  \label{tbl:initval_ringmod}
  \begin{tabular}{rrr|rrr}
    \hline
    $j$ & \multicolumn{1}{c}{$x_j(0)$} & \multicolumn{1}{c|}{$\dot{x}_j(0)$} & \multicolumn{1}{c}{$j$} & \multicolumn{1}{c}{$x_j(0)$} & \multicolumn{1}{c}{$\dot{x}_j(0)$} \\ \hline \hline 
    1 & 0 &                              0 &  9 & 0 & 0 \\
    2 & 0 &                              0 & 10 & 0 & 0 \\
    3 & 0 &  $6.2831853071796 \times 10^4$ & 11 & 0 & 0 \\
    4 & 0 & $-6.2831853071796 \times 10^4$ & 12 & 0 & 0 \\
    5 & 0 & $-6.2831853071796 \times 10^4$ & 13 & 0 & 0 \\
    6 & 0 &  $6.2831853071796 \times 10^4$ & 14 & 0 & 0 \\
    7 & 0 &                              0 & 15 & 0 & 0 \\
    8 & 0 &                              0 &    &   &   \\\hline
  \end{tabular}
\end{table}

\end{document}